\documentstyle[aps,epsf]{revtex}
\renewcommand{\theequation}{\arabic{section}.\arabic{equation}}
\begin{document}
\draft
%
%
%
%
\title{
Exact renormalization group flow equations
for non-relativistic fermions: scaling
towards the Fermi surface}
\author{Peter Kopietz and Tom Busche}
\address{
Institut f\"{u}r Theoretische Physik, Universit\"{a}t Frankfurt,
Robert-Mayer-Strasse 8, 60054 Frankfurt, Germany}
\date{March 30, 2001}
\maketitle
\begin{abstract}
We construct exact functional renormalization group (RG)
flow equations for  
non-relativistic fermions in arbitrary dimensions, 
taking into account not only mode elimination 
but also the rescaling  of the
momenta, frequencies and the fermionic fields.
The complete RG flow of all relevant, marginal and irrelevant couplings 
can be described by
a system of coupled flow equations for the 
irreducible $n$-point vertices. 
Introducing suitable dimensionless variables,
we obtain flow equations for generalized scaling functions 
which are {\it{continuous}} functions of the flow parameter,  
even if we consider quantities 
which are dominated by momenta close to the Fermi surface, such as
the density-density correlation function at long wavelengths.
We also show how the problem of constructing the
renormalized Fermi surface can be reduced to the problem
of finding the RG fixed point of the 
irreducible two-point vertex at vanishing momentum and frequency.
We argue that only if the degrees of freedom are properly
rescaled it is possible to reach 
scale-invariant non-Fermi liquid fixed points
within a truncation of the exact RG flow
equations.
\end{abstract}
\pacs{PACS numbers: 71.10-w, 71.10.Hf}
%
%
%
%
%

\section{Introduction}
\label{sec:intro}

Recently several authors 
\cite{Zanchi96,Halboth99,Halboth00,Honerkamp00,Honerkamp01}
have used exact functional renormalization group (RG) methods to
gain a deeper understanding of strongly correlated 
non-relativistic fermions in reduced dimensions.
The exact functional RG 
yields an infinite hierarchy
of coupled differential equations describing the
change of the correlation functions due to the elimination and
the rescaling of the degrees of freedom.
For classical statistical mechanics problems
the exact functional RG has been developed long time ago
in a pioneering work by Wegner and Houghton \cite{Wegner73}, who
performed the usual three RG-steps \cite{Wilson72,Fisher98}
to derive their exact flow equation:

 \begin{enumerate}

 \item Integrate out fields with momenta ${\bf{k}}$ in a shell
  $\Lambda_0 ( 1 - dt ) < |{\bf{k}} | < \Lambda_0 $,
 where $dt$ is infinitesimal and $\Lambda_0$ is some 
ultraviolet cutoff.
  
 \item
 Rescale the remaining  momenta by a factor $1+ dt$ and express all quantities
in terms of the rescaled momenta.

\item  Rescale the remaining fields by a factor of 
$1 +   \frac{1}{2} ( 1 - \eta ) dt $, where the anomalous dimension $\eta$ has to be chosen such that the RG
has a fixed point \cite{Bell74}.

\end{enumerate}

Over the years several alternative
formulations of the exact functional
RG in field theory and statistical physics have been proposed 
\cite{Nicoll76,Polchinski84,Wetterich93,Morris94,Salmhofer98,Bagnuls00}.
Note, however, that recent applications of the exact RG 
for two-dimensional 
fermions \cite{Zanchi96,Halboth99,Halboth00,Honerkamp00,Honerkamp01}
take only the 
mode elimination step 1 into account, 
and do not perform the above rescaling 
steps 2 and 3.
While such a procedure is legitimate if one 
is interested in solving a given many-body problem 
iteratively,  the Wilsonian RG amounts to 
more than that \cite{Fisher98}:  
by combining the mode elimination
with a suitable rescaling of the degrees of freedom,
the Wilsonian RG generates a mapping of the original many-body problem
onto a continuous family of 
new many-body problems, labeled by a flow parameter $t$.
In the limit $t \rightarrow \infty$ it sometimes happens that
the resulting problem simplifies and can be solved in a controlled  
manner.
In order to obtain a non-trivial RG fixed point for $t \rightarrow \infty$
which is characterized
by a non-zero anomalous dimension $\eta$,
it is crucial that the fields are properly rescaled, as discussed
in detail by Bell and Wilson \cite{Bell74}.
For example,
by applying the exact functional 
RG to $\phi^4$-theory \cite{Morris94,Kopietz01},
it is easy to see that 
without the rescaling the RG flow does not
reproduce the Wilson-Fisher fixed point below four dimensions, 
see Sec.\ref{subsec:wilson}.  
We therefore believe that the
RG calculations of 
Refs.\cite{Zanchi96,Halboth99,Halboth00,Honerkamp00,Honerkamp01}
are incomplete, and should be augmented by the usual rescaling steps 2 and 3
given above.
Note that the rescaling is explicitly included in the
one loop momentum-shell approach advanced by Shankar\cite{Shankar94}, 
so that  one may wonder why the exact functional RG  used in
Refs.\cite{Zanchi96,Halboth99,Halboth00,Honerkamp00,Honerkamp01}
does not include the rescaling. We shall try to clarify this point 
in this work by showing how to
include the rescaling of the degrees of freedom
in the exact functional RG
for fermionic many-body systems in arbitrary dimensions.
We then point out several advantages of working with
the rescaled version of the exact RG flow equations.

\section{Mode elimination for the Legendre effective action}
\label{sec:mode}
\setcounter{equation}{0}

In this section we present a brief derivation of the 
exact flow equation describing the change in the
Legendre effective action
$\Gamma \{ \bar{\varphi} , \varphi \}$ due to the
elimination of the degrees of freedom with momenta and frequencies
in a suitably defined shell.
Keeping in mind that the
Legendre effective action is the generating functional
of the one-particle irreducible correlation functions \cite{ZinnJustin89},
an expansion of $\Gamma \{ \bar{\varphi} , \varphi \}$ in powers
of the fields $\bar{\varphi} $ and $\varphi$ yields  the
corresponding flow equations for the irreducible $n$-point vertices.
For classical field theories, the  flow equation for the
Legendre effective action has  been considered some time ago by
Nicoll et al.\cite{Nicoll76}, and more recently by
Wetterich \cite{Wetterich93}, and by Morris \cite{Morris94}.
For non-relativistic fermions the exact flow equation for the
Legendre effective action has recently been derived in a very general form 
by Honerkamp, Salmhofer and co-authors\cite{Honerkamp00,Honerkamp01}, who also
included the possibility of symmetry breaking.
Our derivation given below is less general but more explicit;
in particular, we shall 
adopt as much as possible the standard notations from the theory
of critical phenomena \cite{Wegner73,Kopietz01}.

\subsection{Assumptions and initial conditions}
\label{subsec:FS}

\begin{figure}
\epsfysize4.0cm 
\hspace{5mm}
\epsfbox{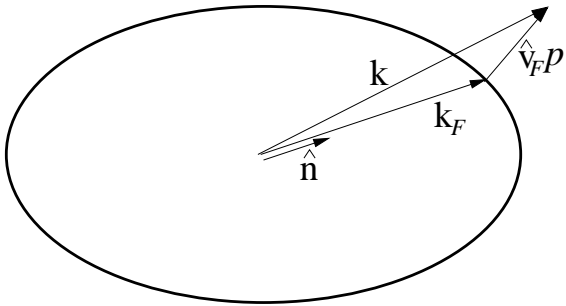}
\vspace{5mm}
\caption{
An arbitrary vector ${\bf{k}}$  
can be decomposed into a component
${\bf{k}}_F$ on the Fermi surface (here an ellipse) and a 
component $\hat{\bf{v}}_F p$
parallel to the local Fermi velocity, see Eq.(\ref{eq:kFdef}).
This construction defines ${\bf{k}}_F$ as a function of ${\bf{k}}$.
}
\label{fig:FS}
\end{figure}

We consider a translationally invariant
non-relativistic system of interacting fermions 
with a Fermi surface.
Throughout this work we assume that the
fermions are in the normal state, so that
the propagator does not have any anomalous components.
By {\it{Fermi surface}} we mean the
true Fermi surface of the interacting system, which is 
defined as the set of all  momenta  satisfying\cite{Luttinger60}
 \begin{equation}
  \epsilon_{{\bf{k}}_F} = \mu - \Sigma ( {\bf{k}}_F , i 0 ) 
 \; ,
 \label{eq:FSdef}
 \end{equation}
where 
$\epsilon_{{\bf{k}}}$ is the energy dispersion
of the non-interacting system,
 $\mu$ is the exact chemical potential
and $\Sigma ( {\bf{k}} , i \omega_n )$ is the exact self-energy 
of the interacting many-body system.
Here ${\bf{k}}$ is the momentum and $i \omega_n $ is the
Matsubara frequency.
Given an arbitrary ${\bf{k}}$, we may define a corresponding  
momentum ${\bf{k}}_F$ on the Fermi surface 
via the decomposition \cite{footnoteFS}
  \begin{equation}
{\bf{k}}   =  {\bf{k}}_F  + \hat{\bf{v}}_F p
 \; ,
 \label{eq:kFdef}
\end{equation}
as shown in Fig.\ref{fig:FS}. Here
$\hat{\bf{v}}_F = {\bf{v}}_F / | {\bf{v}}_F|  $
is a unit vector in the direction of the local Fermi velocity
 \begin{equation}
 {\bf{v}}_F = \nabla_{\bf{k}} \left. \epsilon_{\bf{k}}
 \right|_{ {\bf{k}} = {\bf{k}}_F }
 \; .
 \end{equation}
Note that ${\bf{v}}_F$ is defined in terms of the gradient
of the {\it{bare energy dispersion}} at the {\it{renormalized
Fermi surface}}.
The solution of Eq.(\ref{eq:FSdef}) can be parameterized as
 \begin{equation}
 {\bf{k}}_F = \hat{\bf{n}} k_F ( \hat{\bf{n}} )
 \label{eq:kfsol}
 \; ,
 \end{equation}
where $\hat{\bf{n}}$ is a unit vector in the direction of
${\bf{k}}_F$, see Fig.\ref{fig:FS}.
Note that only 
for a spherical Fermi surface we may identify
$\hat{\bf{n}} = \hat{\bf{v}}_F$, so that 
${\bf{k}}_F$ and  ${\bf{v}}_F$ are  parallel. In this case we know that
${\bf{k}}_F$ is not renormalized as we turn on  the interactions
at constant density \cite{Luttinger60,Blagoev97}.
The true Fermi surface
is then given by $\epsilon_{{\bf{k}}_F} = \mu_0$ (where
$\mu_0$ is the chemical potential of the non-interacting system
at the same density), so that
$\Sigma ( {\bf{k}}_F , i 0 ) = \mu - \mu_0$,
which is independent of ${\bf{k}}_F$.  

For the derivation of the RG flow equations it is convenient
to represent the generating functionals of the
fermionic correlation functions in terms of
Grassmannian functional integrals \cite{Negele88}.
We assume that initially 
the Grassmann fields $\psi_{K}$ with 
large excitation energies 
 $
 \epsilon_{\bf{k}} - \epsilon_{{\bf{k}}_F}$
and  large Matsubara frequencies $\omega_n$
have been integrated out. 
Here $K = ( {\bf{k}} , i \omega_n)$ is a composite label, and for simplicity
we ignore the spin degree of freedom.
The high energy shell can be
described by an equation of the form
$ \Omega_K 
\raisebox{-0.5ex}{$\; \stackrel{>}{\sim} \;$} \xi_0$, where
$\Omega_K$ is a suitable homogeneous function of
$ \epsilon_{\bf{k}} - \epsilon_{{\bf{k}}_F}$
or $ | \omega_n |$.
Possible choices are $ \Omega_K = | \omega_n |$,
$ \Omega_K = 
|  \epsilon_{\bf{k}} - \epsilon_{ {\bf{k}}_F }|$, 
or $\Omega_K = \sqrt{ \omega_n^2 + 
( \epsilon_{\bf{k}} - \epsilon_{ {\bf{k}}_F })^2
  }$.
Our starting point is 
an effective action  of the form
 \begin{eqnarray}
 S_{\xi_0} \{ \bar{\psi}, \psi \} & = & S^0_{\xi_0} \{ \bar{\psi}, \psi \} + 
 S^{\rm int}_{\xi_0} \{ \bar{\psi} , \psi \}
 \;  , 
 \end{eqnarray}
where the free part is given by
 \begin{eqnarray}
 S^0_{\xi_0} \{ \bar{\psi} , \psi \} 
 = \int_{ K } \Theta_{\gamma} ( \xi_0 - 
 \Omega_K )
 [ - i \omega_n + \epsilon_{\bf{k}} - 
 \epsilon_{ {\bf{k}}_F } ] \bar{\psi}_K \psi_{K} 
 \; .
 \nonumber
 \\
 & &
 \label{eq:S0def}
 \end{eqnarray}
Here we use the notation
 \begin{eqnarray}
 \int_K & = & \frac{1}{\beta V} \sum_{ {\bf{k}} , \omega_n }
 \rightarrow \int \frac{ d {\bf{k}} }{ ( 2 \pi )^D }
 \int_{ - \infty}^{\infty} \frac{ d \omega}{ 2 \pi }
 \; ,
 \end{eqnarray}
where $\beta$ is the inverse temperature and $V$ is the volume, and
the right-hand side is valid for $\beta \rightarrow \infty$ and
$V \rightarrow \infty$.
The interaction part is 
 \begin{eqnarray}
  S^{\rm int}_{\xi_0} \{ \bar{\psi} , \psi \} 
& = &
 \nonumber
 \\
 & & \hspace{-15mm} 
 \int_{ K }
  \Theta_{\gamma} ( \xi_0 - \Omega_K) 
 [ \Sigma_{\xi_0} (K)  - 
 \Sigma ( {\bf{k}}_F , i0) ] \bar{\psi}_{K}  \psi_{K} 
 \nonumber
 \\
&  & \hspace{-18mm}
 + \frac{1}{ (2 !)^2} \int_{K_1^{\prime}}
 \int_{ K_2^{\prime} } \int_{K_2} \int_{K_1}
 \delta_{ K_1^{\prime} + K_2^{\prime} , K_2 + K_1}
 \nonumber
 \\
 & & \hspace{-10mm} \times
 \Gamma_{\xi_0}^{(4)} ( K_1^{\prime} , K_2^{\prime} ; K_2 , K_1 )
 \bar{\psi}_{ K_1^{\prime} } 
  \bar{\psi}_{ K_2^{\prime} } 
  \psi_{K_2} \psi_{K_1}
 + \ldots
  \; ,
 \label{eq:Sint}
 \end{eqnarray}
where the ellipsis denotes three-body and higher order interactions, and
 \begin{eqnarray}
 \delta_{ K  K^{\prime} } & = & \beta V 
 \delta_{ {\bf{k}}  {\bf{k}}^{\prime} }
 \delta_{ \omega_n  \omega_{n^{\prime}} }
 \nonumber
 \\
 & \rightarrow & ( 2 \pi )^{D+1} \delta ( 
 {\bf{k}} - {\bf{k}}^{\prime}  ) \delta ( \omega - \omega^{\prime} )
 \; .
 \end{eqnarray}
All vertices are antisymmetric with respect to the
permutation of any pair of the incoming
particles 
and any pair of the outgoing particles.
In particular, the four-point
vertex $\Gamma_{\xi}^{(4)} ( K_1^{\prime} , K_2^{\prime} ; K_2 , K_1 )$
is antisymmetric with respect to the 
exchange $K_1^{\prime} \leftrightarrow K_2^{\prime}$ and 
 $K_1 \leftrightarrow K_2$.
The function
$\Theta_{\gamma} (  \epsilon )$ is a smooth cutoff function, satisfying
$\Theta_{\gamma} ( \epsilon ) \approx 1$ for $   \epsilon > \gamma$ and
$\Theta_{\gamma} ( \epsilon ) \approx 0$ for $ \epsilon  < - \gamma$, so that
$\lim_{\gamma \rightarrow 0} \Theta_{\gamma} ( \epsilon ) = 
\Theta ( \epsilon )$.
The term $\Sigma_{\xi_0} (K)$ in Eq.(\ref{eq:Sint}) is
the contribution from the high energy fields with 
$ \Omega_K  \raisebox{-0.5ex}{$\; \stackrel{>}{\sim} \;$}
 \xi_0$ 
to the irreducible self-energy, and 
$ - \Sigma ( {\bf{k}}_F , i0)$ is a counterterm which 
takes into account that
in the free action (\ref{eq:S0def}) we have subtracted 
$ \epsilon_{{\bf{k}}_F} = \mu - \Sigma ( {\bf{k}}_F , i 0 )$
from the bare energy dispersion.
As shown below, this subtraction 
is crucial to obtain the RG flow of the
Fermi surface.
In a perturbative approach, the above subtraction
is necessary to obtain a well-behaved perturbation series
\cite{Luttinger60,Metzner98}. 
The counterterm 
$ - \Sigma ( {\bf{k}}_F , i0)$ can be determined 
from a self-consistency condition, which can be
imposed order by order in perturbation theory \cite{Feldmann96}.
Within a RG approach, the counterterm can be determined
a posteriori
from the condition that the flow equation of the 
momentum- and frequency-independent part of the 
irreducible two-point vertex has a fixed 
point \cite{Shankar94}, see  Sec.\ref{sec:FSshape} for a careful
discussion. 
The energy scale $\xi_0$  
plays a dual role \cite{Morris94}:
For the calculation of 
the vertices of the initial action 
$ S_{\xi_0} \{ \bar{\psi}, \psi \}$
the scale $\xi_0$ acts as an {\it{infrared}} cutoff, so that
for sufficiently large $\xi_0$ the 
vertices appearing in
$ S_{\xi_0} \{ \bar{\psi} , \psi \}$
can be calculated perturbatively.
On the other hand, for the remaining low-energy degrees of
freedom $\xi_0$ plays the role of an
{\it{ultraviolet}} cutoff.

\subsection{Exact flow equation describing mode elimination}

Starting from  the effective action $S_{\xi_0} \{ \bar{\psi} , \psi \}$
with cutoff $\xi_0$,
we eliminate all 
fields with momenta and Matsubara frequencies in the regime  
$\xi 
\raisebox{-0.5ex}{$\; \stackrel{<}{\sim} \;$}
 \Omega_K 
\raisebox{-0.5ex}{$\; \stackrel{<}{\sim} \;$} \xi_0$.  
The free propagator
for the fermionic fields in this regime 
has in the $K$-basis the diagonal elements
 \begin{equation}
  G^0_{\xi , \xi_0}  ( K )  =
 \frac{ \Theta_{\gamma} (  \Omega_K  - 
 \xi )  -  \Theta_{\gamma} (  \Omega_K  -  \xi_0 )
}{ i \omega_n - \epsilon_{\bf{k}}  + \epsilon_{ {\bf{k}}_F } }
 \; .
 \label{eq:propdef}
 \end{equation}
Because ${\bf{k}}_F $ lies by construction on the true 
Fermi surface of the {\it{interacting}} system,
the right-hand side of Eq.(\ref{eq:propdef}) depends implicitly
on the interaction.
Note, however, that $\epsilon_{ {\bf{k}}_F}$ 
is independent of the flow parameter $\xi$.
Diagrammatically, the 
elimination of the degrees of freedom in the shell
$\xi 
\raisebox{-0.5ex}{$\; \stackrel{<}{\sim} \;$}
 \Omega_K 
\raisebox{-0.5ex}{$\; \stackrel{<}{\sim} \;$} \xi_0$
corresponds to contracting all terms  generated by expanding
$e^{-S_{\xi_0}^{\rm int} \{ \bar{\psi} , \psi \}}$ with the propagators
$G^0_{\xi , \xi_0}$ given in Eq.(\ref{eq:propdef}).
The connected correlation functions of the new theory 
can be formally represented as functional derivatives
of the generating functional 
${\cal{G}}^c_{\xi} \{ 
 \bar{J}  , {J}  \}$ defined by \cite{Ellwanger94}
 \begin{eqnarray}
 e^{ {\cal{G}}^c_{\xi} \{ 
 \bar{J}  , {J}  \} }&  = & e^{ - S^{\rm int}_{\xi_0} \{ 
 \zeta \frac{\delta}{ \delta J} , \frac{ \delta }{\delta \bar{J}} \} }
 e^{ ( \bar{J} , ( - G^0_{\xi , \xi_0} ) J ) }
 \nonumber
 \\
 &  & \hspace{-16mm} 
 = e^{  ( \bar{J} , ( - G^0_{\xi , \xi_0} ) J ) }
 \left[ e^{ ( \zeta \frac{\delta}{\delta \psi} , ( - G^0_{\xi , \xi_0} )
 \frac{\delta}{\delta \bar{\psi}} ) } 
 e^{ - S^{\rm int}_{\xi_0} \{ \bar{\psi} , \psi \} } 
 \right]_{ \psi = G^0_{\xi , \xi_0} J }
 \; .
 \nonumber
 \\
 & &
 \label{eq:Gcdef}
 \end{eqnarray} 
Here  $( a , M b ) = \int_K \int_{K^{\prime}}
a_K M_{K  K^{\prime}} b_{K^{\prime}}$,
and the factor $\zeta = -1$ arises from the antisymmetry of the
Grassmann fields \cite{Negele88,footnotegrass}. 
$G^0_{\xi , \xi_0}$ should be considered as a matrix in $K$-space,
with matrix elements given by
 \begin{equation}
 [ G^0_{\xi , \xi_0} ]_{ K K^{\prime} } = \delta_{K  K^{\prime}}
 G^0_{\xi , \xi_0} ( K )
 \; ,
 \end{equation}
where $G^0_{\xi , \xi_0} ( K )$ is defined in Eq.(\ref{eq:propdef}).

Differentiating both sides of Eq.(\ref{eq:Gcdef})
with respect to $\xi$, we obtain an exact flow equation
for the generating functional of the
connected correlation functions.
For practical calculations the flow 
equation for the irreducible vertices is
more convenient \cite{Nicoll76,Wetterich93,Morris94}. 
To obtain the corresponding generating functional,
we perform a Legendre transformation,
 \begin{equation}
 {\cal{L}}_{\xi} \{ \bar{\varphi} , \varphi \}  = ( \bar{\varphi} , J )
 + ( \bar{J} , \varphi ) - {\cal{G}}^c_{\xi} \{ 
 \bar{J}  , {J}  \}
 \; ,
 \end{equation}
where the Grassmann sources $J$ and $\bar{J}$
have to be considered as functionals of the 
Grassmann fields $\varphi$ and $\bar{\varphi}$ by solving
the following equations
for
 $ \bar{J} = \bar{J} \{ \bar{\varphi} , \varphi \} $ and $J =
 {J} \{ \bar{\varphi} , \varphi \}$, 
 \begin{equation}
 \varphi = \zeta \frac{ \delta {\cal{G}}^c_{\xi} \{ \bar{J} , J \} }{
 \delta J } \; \;  , \; \; 
 \bar{\varphi} =  \frac{ \delta {\cal{G}}^c_{\xi} \{ \bar{J} , J \} }{
 \delta \bar{J} } 
 \; .
 \end{equation} 
The generating functional of the irreducible vertices is
then given by 
\begin{equation}
 \Gamma_{\xi}  \{ \bar{\varphi} , \varphi \}
 =   {\cal{L}}_{\xi} \{ \bar{\varphi} , \varphi \} 
 - ( \bar{\varphi} , [ - G^0_{\xi , \xi_0 } ]^{-1} \varphi )
 \; ,
 \end{equation}
and satisfies the exact flow 
equation \cite{Honerkamp00,Honerkamp01,footnoteflow}
 \begin{eqnarray}
 \partial_{\xi} \Gamma_{\xi}
 & = &  \zeta \int_K  
  \partial_{\xi} [ G^0_{\xi ,\xi_0 } (K) ]^{-1} 
 \nonumber
 \\
 & \times & 
  \left\{
  \hat{G}_{\xi , \xi_0 }  \hat{\cal{U}}_{\xi} 
 [ \hat{1} - \hat{G}_{\xi , \xi_0 }
 \hat{\cal{U}}_{\xi} ]^{-1} \hat{G}_{\xi , \xi_0} 
 \right.
 \nonumber \\
 &  &  \left.
 + \hat{G}^0_{\xi , \xi_0 } \hat{\Sigma}_{\xi} 
[ 1 - \hat{G}^0_{\xi , \xi_0} \hat{\Sigma}_{\xi} ]^{-1} 
 \hat{G}^0_{\xi , \xi_0}  
 \right\}_{22;KK} \; ,
 \label{eq:flowgeneral}
 \end{eqnarray}
with initial condition
 \begin{equation}
 \Gamma_{\xi_0} \{ \bar{\varphi} , \varphi \} =
 S^{\rm int}_{\xi_0} \{ \bar{\varphi} , \varphi \}
 \; .
 \end{equation}
Here $\{ \ldots \}_{22;KK}$ denotes the lower diagonal element of the
corresponding $2 \times 2$-matrix in $K$-space, and
the functional $\hat{\cal{U}}_{\xi}$ is given  by
\begin{equation}
 \left( 
 \begin{array}{cc}
 \zeta \frac{ \delta^2 \Gamma_{\xi} }{
 \delta \bar{\varphi}_K \delta \varphi_{K^{\prime}} } & 
 \zeta \frac{ \delta^2 \Gamma_{\xi} }{ \delta \bar{\varphi}_K
  \delta \bar{\varphi}_{K^{\prime}}} \\
   \frac{ \delta^2 {\Gamma}_{\xi} }{
 \delta {\varphi}_{K} \delta \varphi_{K^{\prime}} } & 
  \frac{ \delta^2 \Gamma_{\xi} }{ \delta {\varphi}_{K}
  \delta \bar{\varphi}_{K^{\prime}}}
 \end{array} 
 \right) 
 = [ \hat\Sigma_{\xi} ]_{K  K^{\prime} } + 
\left[ \hat{\cal{U}}_{\xi} \{ \bar{\varphi } 
 , \varphi \} \right]_{K  K^{\prime}} 
 \; ,
\end{equation}
where $\hat{\Sigma}_{\xi}$ is defined as the field-independent part
of the second functional derivative of $\Gamma_{\xi}$, so that
$ \hat{\cal{U}}_{\xi} \{ 0 , 0 \} = 0$.
The interacting cutoff-regularized propagator
is related to the non-interacting one via the Dyson equation,
 \begin{equation}
 \hat{G}_{\xi, \xi_0 } = [ \hat{G}^0_{\xi ,\xi_0 } - \hat{\Sigma}_{\xi} ]^{-1}
 \; \; , \; \; 
 [ \hat{G}^0_{\xi, \xi_0} ]_{ij} = \delta_{ij} 
 G^0_{\xi , \xi_0} 
 \; .
 \end{equation}

\subsection{Flow equation for sharp cutoff}
\label{subsec:sharp} 

For simplicity we shall from now on work with a sharp cutoff function,
$\Theta ( \epsilon ) = \lim_{\gamma \rightarrow 0} \Theta_{\gamma} 
( \epsilon )$.
Due to translational invariance
the $K$-dependence of all vertices is constrained
by energy-momentum conservation, so that
the matrix elements of
 $\hat{\Sigma}_{\xi}$ are
 \begin{eqnarray}
  [ \hat{\Sigma}_{\xi} ]_{ij ; K K^{\prime}} &  = & 
 \delta_{ij}  \delta_{K K^{\prime}} [ \Sigma_{\xi} ( K )  - \Sigma ( {\bf{k}}_F , i 0 ) ]
 \; ,
 \label{eq:sigmadiag}
 \end{eqnarray}
where the term $\Sigma ( {\bf{k}}_F , i 0 )$ 
is due to the subtraction in Eq.(\ref{eq:Sint}).
The exact flow equation (\ref{eq:flowgeneral}) can then be 
reduced to the following form \cite{Morris94,Kopietz01} 
 \begin{eqnarray}
 \partial_{\xi} \Gamma_{\xi}
 & = &  
 \nonumber
 \\
 & & \hspace{-10mm}
\zeta \int_K \frac{ \delta ( \Omega_K - \xi ) }{
 i \omega_n - \epsilon_{\bf{k}} + \mu - \Sigma_{\xi} ( K ) }
 \left\{
  \hat{\cal{U}}_{\xi} [ \hat{1} - \hat{G}_{\xi , \xi_0}
 \hat{\cal{U}}_{\xi} ]^{-1}  \right\}_{22;KK}
 \nonumber
 \\
 &  &  \hspace{-10mm} - \zeta  \beta V  \int_K \delta ( \Omega_K - \xi )
 \ln \left[ \frac{ i \omega_n - \epsilon_{\bf{k}} + \mu
 - \Sigma_{\xi} ( K ) }{  i \omega_n - \epsilon_{\bf{k}} + 
 \epsilon_{ {\bf{k}}_F} }
 \right]
 \; .
 \nonumber
 \\
 & &
 \label{eq:flowsharp}
 \end{eqnarray} 
We now expand
 \begin{eqnarray}
 \Gamma_{\xi} \{ \bar{\varphi} , \varphi \} & = & 
 \sum_{n = 0}^{\infty} \frac{ (-1 )^n}{ ( n ! )^2 }
 \int_{ K_1^{\prime} } \ldots \int_{ K_n^{\prime} }
 \int_{ K_n } \ldots \int_{ K_1 }
 \nonumber
  \\
 &   \times &
 \delta_{ K_1^{\prime} + \ldots + K_n^{\prime} , K_n + \ldots + K_1 }
 \nonumber
 \\
 & \times &
 \Gamma_{\xi}^{(2n)}  ( K_1^{\prime}, \ldots ,
 K_n^{\prime} ; K_n , \ldots , K_1 )
 \nonumber
 \\
 & \times & 
 \bar{\varphi}_{K_1^{\prime}} \cdots \bar{\varphi}_{ K_n^{\prime}}
 {\varphi}_{K_n} \cdots {\varphi}_{ K_1}
 \; ,
 \label{eq:Gammaexpand}
 \end{eqnarray}
and identify 
the terms with the same powers of the fields
on both sides of Eq.(\ref{eq:flowsharp}).
Note that in the normal state only the even vertices are non-zero.
In this way we obtain
the RG flow equations for the unrescaled irreducible
$2n$-point vertices $
 \Gamma_{\xi}^{(2n)}  ( K_1^{\prime}, \ldots ,
 K_n^{\prime} ; K_n , \ldots , K_1 )$.
We now explicitly give the exact flow equations 
for the vertices
$\Gamma^{(0)}_{\xi}$,
$\Gamma^{(2)}_{\xi}$, and
$\Gamma^{(4)}_{\xi}$.
For a two-loop calculation one needs the flow equation
for the six-point vertex $\Gamma^{(6)}_{\xi}$,
which is given in the Appendix.

\subsubsection{Free energy}
The interaction correction to the free energy is obtained
from the last term in Eq.(\ref{eq:flowsharp}),
 \begin{equation}
 \partial_{\xi} \Gamma^{(0)}_{\xi} =
 - \zeta  \beta V  \int_K \delta ( \Omega_K - \xi )
 \ln \left[ \frac{ i \omega_n - \epsilon_{\bf{k}} + \mu
 - \Sigma_{\xi} ( K ) }{  i \omega_n - \epsilon_{\bf{k}} 
 + \epsilon_{ {\bf{k}}_F } }
 \right]
 \; .
 \label{eq:flowGamma0}
 \end{equation}

\subsubsection{Self-energy}
\begin{figure}
\epsfysize2.5cm 
\hspace{5mm}
\epsfbox{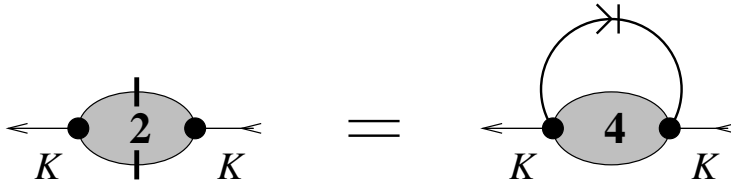}
\vspace{5mm}
\caption{
Diagrammatic representation of the flow equation
for the two-point vertex, see Eqs.(\ref{eq:flowGamma2})
and (\ref{eq:twopointscale}).
The left-hand side represents the derivative
of the total irreducible vertex with respect to the flow
parameter.
The arrows represent the exact propagators, and an arrow with an extra
slash  represents the derivative of the cutoff function
with respect to the flow parameter.
}
\label{fig:TwoPoint}
\end{figure}

Comparing the terms quadratic in the fields on both sides of
Eq.(\ref{eq:flowsharp}) and using the fact that
by construction 
the two-point vertex is related to  the
subtracted irreducible self-energy via 
 \begin{equation}
 \Gamma_{\xi}^{(2)} ( K; K )
  = - [   \Sigma_{\xi} ( K )  - \Sigma ( {\bf{k}}_F , i 0 ) ]
 \; ,
 \end{equation}
we obtain
 \begin{eqnarray}
 \partial_\xi \Gamma^{(2)}_{\xi} ( K ; K ) & = &
 - \partial_{\xi} \Sigma_{\xi} ( K ) = 
 \nonumber
 \\
 & & \hspace{-25mm}
- \zeta \int_{ K^{\prime}}
 \frac{ \delta ( \Omega_{K^{\prime}} - \xi )}{
 i \omega_{n^{\prime}} - \epsilon_{ {\bf{k}}^{\prime} } + \mu
 - \Sigma_{\xi} ( K^{\prime} ) }
 \Gamma^{(4)}_{\xi} ( K , K^{\prime} ; K^{\prime} , K )
 \; .
 \nonumber
 \\
 & &
 \label{eq:flowGamma2}
 \end{eqnarray} 
A graphical representation of this equation is shown in
Fig.\ref{fig:TwoPoint}.

\subsubsection{Four-point vertex}
\begin{figure}
\epsfysize8cm 
\hspace{5mm}
\epsfbox{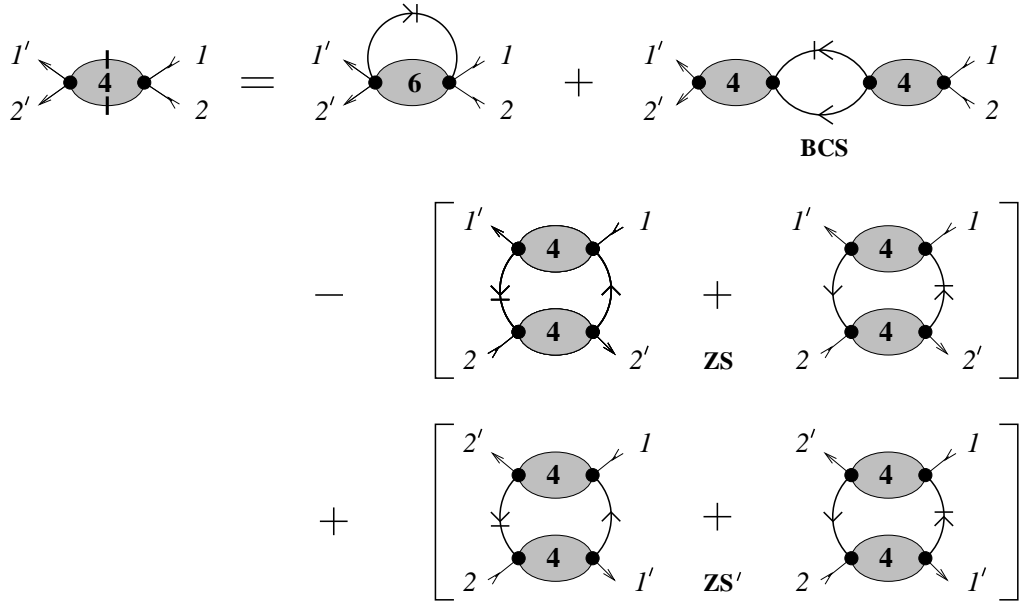}
\vspace{5mm}
\caption{
Diagrammatic representation of the flow equation
for the four-point vertex, see Eqs.(\ref{eq:flowGamma4}) and
(\ref{eq:fourpointscale}).
}
\label{fig:FourPoint}
\end{figure}

The flow equation for the irreducible four-point vertex is
 \begin{eqnarray}
 \partial_\xi \Gamma^{(4)}_{\xi} ( K_1^{\prime} , K_2^{\prime} ; K_2 , K_1 ) & =&
 \nonumber
 \\
 & & \hspace{-40mm}
  - \zeta \int_{ K}
 \frac{ \delta ( \Omega_K - \xi )}{
 i \omega_{n} - \epsilon_{ {\bf{k}} } + \mu
 - \Sigma_{\xi} ( K ) }
 \Gamma^{(6)}_{\xi} ( K_1^{\prime} , K_2^{\prime} , K  ; K , K_2 , K_1 )
 \nonumber
 \\
 &  & \hspace{-40mm} + \int_K
 \left[  \frac{ \delta ( \Omega_K - \xi )
 G_{\xi , \xi_0} ( K^{\prime} )}{
 i \omega_{n} - \epsilon_{ {\bf{k}} } + \mu
 - \Sigma_{\xi} ( K ) }
 +
 \frac{  G_{\xi , \xi_0} ( K )  
 \delta ( \Omega_{K^{\prime}} - \xi )}{
 i \omega_{n^{\prime}} - \epsilon_{ {\bf{k}}^{\prime} } + \mu
 - \Sigma_{\xi} ( K^{\prime} ) } \right]
 \nonumber
 \\
 & & \hspace{-38mm} \times
\left\{
 \frac{1}{2} \left[ 
 \Gamma^{(4)}_{\xi} ( K_1^{\prime} , K_2^{\prime} ; K^{\prime} , K ) 
 \Gamma^{(4)}_{\xi} ( K , K^{\prime} , K_2 , K_1 ) 
 \right]_{K^{\prime} = K_1 + K_2 - K }
 \right.
 \nonumber
 \\
 & & \hspace{-32mm} + \zeta
 \left[ 
 \Gamma^{(4)}_{\xi} ( K_1^{\prime} ,  K^{\prime} ;  K , K_1 ) 
 \Gamma^{(4)}_{\xi} ( K_2^{\prime} , K ; K^{\prime} , K_2 ) 
 \right]_{K^{\prime} = K + K_1 - K_1^{\prime} }
 \nonumber
 \\
 & & \hspace{-32mm}  \left. +
  \left[ 
 \Gamma^{(4)}_{\xi} ( K_2^{\prime} ,  K^{\prime} ;  K , K_1 ) 
 \Gamma^{(4)}_{\xi} ( K_1^{\prime} , K ; K^{\prime} , K_2 ) 
 \right]_{K^{\prime} = K + K_1 - K_2^{\prime} }
 \right\}
 \; .
 \label{eq:flowGamma4}
 \end{eqnarray} 
This equation is shown graphically in Fig.\ref{fig:FourPoint}.
The first term in the curly braces is contribution from the
BCS-channel, 
while the last two terms are the zero-sound 
contributions, usually abbreviated by ZS (second term)
and ZS$^{\prime}$ (third term) \cite{Shankar94}.
Note that both zero-sound terms have to be retained in order
to preserve the antisymmetry of the four-point 
vertex \cite{Chitov95}. 
The exact flow equation for the six-point vertex is 
rather lengthy and is given in the Appendix.

\section{Exact flow equations describing mode elimination and rescaling}
\label{sec:rescale}
\setcounter{equation}{0}

So far we have derived exact flow equations 
for the irreducible vertices describing the
elimination of the degrees of freedom.
Within the one-loop approximation,
it is sufficient to
set $\Gamma^{(2n)}_{\xi} = 0$ for $n \geq 3$,
and to ignore interaction corrections to the propagators
in internal loops. The resulting truncated flow equation
for the four-point vertex in two dimensions 
has been analyzed numerically
by Honerkamp et al.\cite{Honerkamp00,Honerkamp01}.
A similar numerical analysis of an equivalent  
truncated flow equation has been performed by Halboth 
and Metzner \cite{Halboth99,Halboth00}.
Both groups found that the one-loop flow of the four-point vertex
eventually diverges at a finite scale, where the perturbative RG
breaks down. Physically the runaway flow has been interpreted
in terms of some incipient instability of the normal metallic phase.
Here we would like to point out that in principle
there is another interpretation of this runaway flow, namely
the existence of a scale-invariant non-Fermi liquid fixed point,
which is characterized by a finite anomalous dimension.
Below we argue that the mode elimination RG transformations  
of Refs.\cite{Zanchi96,Halboth99,Halboth00,Honerkamp00,Honerkamp01}
cannot detect such a fixed point, 
because these equations do not take into account that
the degrees of freedom should be properly rescaled 
to reach a fixed point with a finite anomalous dimension \cite{Bell74}. 
Assuming the existence of such a fixed point,
only the rescaled version of the exact flow equations
given below would detect it, while the
pure mode elimination RG 
used in Refs.\cite{Zanchi96,Halboth99,Halboth00,Honerkamp00,Honerkamp01}
would still exhibit a runaway flow to strong coupling.

Because the following rescaling procedure 
depends crucially on the existence of a Fermi surface, we now
explicitly set the statistics factor $\zeta = -1$.
We also take the limits of infinite volume ($V \rightarrow \infty$) 
and zero temperature ($\beta \rightarrow \infty$), so that
momenta and frequencies become continuous variables.

\subsection{Scaling variables}
\label{subsec:scaling}

Instead of the momentum ${\bf{k}}$ and the Matsubara frequency
$ i \omega$, we now  label
the degrees of freedom by the 
direction $\hat{\bf{n}}$ of ${\bf{k}}_F$ (see Fig.\ref{fig:FS}) 
and by the
dimensionless variables
 \begin{eqnarray}
 q = \frac{ v_F p}{\xi} & = & \frac{ {\bf{v}}_F \cdot ( {\bf{k}}
 - {\bf{k}}_F )}{\xi} 
 \; \; , \; \; 
\epsilon  = \frac{\omega }{ \xi} 
 \; ,
\end{eqnarray}
so that
 \begin{equation}
 {\bf{k}} = {\bf{k}}_F + \hat{\bf{v}}_F \frac{ \xi  }{v_F}  q
 = \hat{\bf{n}} k_F ( \hat{\bf{n}} ) + \hat{\bf{v}}_F \frac{ \xi  }{v_F}  q
 \; . 
\label{eq:trans}
 \end{equation}
Furthermore, instead of the flow parameter $\xi$, 
we introduce the dimensionless logarithmic flow parameter
 \begin{equation}
 t = - \ln ( \xi / \xi_0 )
 \; ,
 \label{eq:xidef}
 \end{equation}
and define the dimensionless dispersion
 \begin{equation} 
\xi_t^{ \hat{\bf{n}} } ( q  ) = 
\frac{ \epsilon_{ {\bf{k}}  } - 
\epsilon_{ {\bf{k}}_F }}{\xi}
 = q  + \frac{c_t^{ \hat{\bf{n}} }}{2}  q^2 
  + O (q^3 )
 \; ,
 \label{eq:xitqdef}
 \end{equation}
where
 \begin{equation} 
 c_t^{\hat{\bf{n}}} = \frac{\xi}{m v_F^2}  = \frac{ \xi_0}{m v_F^2 } e^{-t}
 \label{eq:ctdef}
 \end{equation}
is an irrelevant coupling which measures the 
leading deviation from linearity in the energy dispersion
in the direction normal to the Fermi surface.
In Eq.(\ref{eq:xitqdef}) we have expanded
the energy dispersion
around ${\bf{k}} = {\bf{k}}_F$,
 \begin{equation}
 \epsilon_{ {\bf{k}} } =  \epsilon_{ {\bf{k}}_F } +
{\bf{v}}_F  \cdot ( {\bf{k}} - {\bf{k}}_F ) + 
\frac{ ( {\bf{k}} - {\bf{k}}_F )^2 }{2 m }
 + \ldots
 \; .
 \end{equation}
Note that in general ${\bf{v}}_F$ and 
$m$  depend on $\hat{\bf{n}}$.
Using $  Q  = 
( \hat{\bf{n}} , q, i \epsilon ) $ instead of
$ K = ( {\bf{k}} , i \omega )$ as integration variables, we 
have for $\beta \rightarrow \infty$ and $V \rightarrow \infty$
\begin{equation}
 \int_K =
\nu_0 \xi^2  \int_Q
= \nu_0 \xi^2 
\int \frac{ d S_{\hat{\bf{n}} }}{S_D}
\int dq J ({\hat{\bf{n}}} , q ) \int \frac{ d \epsilon}{2 \pi}
\label{eq:measure}
\; \; ,
\end{equation}
where $dS_{\hat{\bf{n}}}$ is a surface element and $S_D$
is the surface area of
the unit sphere in $D$ dimensions,
and
 $J ( { \hat{\bf{n}} } ,  q )$ is a dimensionless Jacobian
associated with the transformation ${\bf{k}} \rightarrow 
( \hat{\bf{n}} , q )$. For convenience we have pulled out a factor of
$\nu_0 \xi^2$ in Eq.(\ref{eq:measure}), where
$\nu_0$ is the  density of states at the Fermi surface,  
 \begin{equation}
 \nu_0 = \int \frac{d {\bf{k}}}{ ( 2 \pi )^D} 
 \delta ( \epsilon_{\bf{k}} - \epsilon_{{\bf{k}}_F } )
 \; .
 \end{equation}
With this normalization
$ J ( \hat{\bf{n}} , q )$ is dimensionless.
In particular,
for a spherical Fermi surface
\begin{equation}
\nu_0 = \frac{\Omega_D}{( 2 \pi )^D} \frac{k_F^{D-1}}{v_F}
 \; \; , \; \, 
 J ( \hat{\bf{n}} ,  q ) = ( 1 + c_t q )^{D-1} \; .
\end{equation}

\subsection{The scaling form of the irreducible vertices}

The proper definition of the dimensionless scaling form
of the irreducible  vertices 
follows partially from dimensional analysis, and partially from
aesthetic considerations (such as
the requirement that numerical prefactors should be as simple as possible).
Given the
expansion (\ref{eq:Gammaexpand}) 
of the generating functional of the irreducible
vertices in powers of the dimensionful fields $\varphi_K$, 
we substitute
 \begin{equation}
 \varphi_K = \left( \frac{ Z_t^{\hat{\bf{n}}} }{ \xi^3 \nu_0 } \right)^{1/2}
 \tilde{\varphi}_Q
 \; ,
 \label{eq:fieldscale}
 \end{equation}
where the
wave-function renormalization factor 
$Z_t^{\hat{\bf{n}}}$ is related to the
irreducible self-energy $\Sigma_{\xi} ( {\bf{k}} , i \omega )$
at scale $\xi$ as usual,
\begin{equation}
Z_t^{\hat{\bf{n}}} = \frac{1}{1 - \left. \frac{\partial 
\Sigma_{\xi} ( {\bf{k}}_F , i \omega )}{\partial
( i \omega ) } \right|_{\omega=0}}
\; .
\label{eq:Zdef}
\end{equation}
The dimensionless fields
$\tilde{\varphi}_{Q}$ should be considered as functions
of the scaling variables $ Q $.

\subsubsection{Free energy}
Due to the Jacobian
associated with the rescaling of fermionic fields in the
functional integral, the field-independent part
$\Gamma_{\xi}^{(0)}$ of the generating functional
$\Gamma_{\xi} \{ \bar{\varphi} , \varphi \}$ picks up
an additive term, so that after rescaling
the correction to the free energy is
\begin{equation}
\tilde{\Gamma}^{(0)}_{t} = \Gamma_{\xi}^{(0)} +  \beta V \int_K 
\Theta ( \xi - \Omega_K ) \ln Z_t^{\hat{\bf{n}}}
 \; .
\label{eq:Gamma0t}  
\end{equation} 
In classical statistical mechanics 
a contribution analogous to the second term 
has been discussed by 
Wegner and Houghton \cite{Wegner73}.

\subsubsection{Two-point vertex}
We write the exact propagator in the following scaling form,
  \begin{equation}
 G_{\xi , \xi_0} (  {\bf{k}} , i \omega  )
 = \frac{Z_t^{\hat{\bf{n}}} }{\xi}  
 \tilde{G}_t  \left( \frac{ {\bf{k}}_F }{k_F} , 
 \frac{ {\bf{v}}_F \cdot ( {\bf{k}} - {\bf{k}}_F 
 ) }{\xi}  , \frac{ i \omega}{\xi} \right)
 \; ,
 \label{eq:Gtscale}
 \end{equation}
where $\hat{\bf{n}}$ should be considered  as a function of ${\bf{k}}$, as
given in Eqs.(\ref{eq:kFdef}) and (\ref{eq:kfsol}).
Note that our exact RG equations describe the flow of the
{\it{dimensionless scaling function}}
  \begin{equation}
 \tilde{G}_t ( Q ) 
 \equiv \tilde{G}_t ( {\hat{\bf{n}}} ,  q , i \epsilon ) 
=  \frac{\xi}{Z_t^{\hat{\bf{n}} } } 
 G_{\xi , \xi_0} (  \hat{\bf{n}} k_F ( \hat{\bf{n}} ) 
 + \hat{\bf{v}}_F \frac{\xi}{v_F} q , 
 i \xi \epsilon )
 \; .
 \label{eq:Gtdef}
 \end{equation}
Introducing the dimensionless scaling form of the (subtracted)
irreducible two-point vertex,
 \begin{equation}
  \tilde{\Gamma}_{t}^{(2)}
 ( Q  )  = \frac{Z_t^{\hat{\bf{n}} } }{\xi} \Gamma^{(2)}_{\xi} ( K ) =
 - \frac{Z_t^{\hat{\bf{n}} } }{\xi} 
 \left[ \Sigma_{\xi} ( K ) - \Sigma ( {\bf{k}}_F , i0 ) \right]
 \; ,
 \label{eq:Gammatdef}
 \end{equation}
and the scaling form of the 
inverse propagator
 \begin{eqnarray}
  r_t ( Q ) & = & 
 \frac{Z_t^{\hat{\bf{n}}} }{\xi} \left[ i \omega - \epsilon_{\bf{k}} + \mu
 - \Sigma_{\xi} ( {\bf{k}} , i \omega ) \right]  
 \nonumber
 \\
 & = & 
  Z_t^{\hat{\bf{n}}}  \left[ i \epsilon - 
 \xi_t^{ \hat{\bf{n}}} ( q ) \right] + 
 \tilde{\Gamma}_t^{(2)} ( Q )
 \label{eq:rtdef}
 \; ,
 \end{eqnarray}
we obtain
\begin{eqnarray}
\tilde{G}_t ( Q ) & = & \frac{  
 \Theta ( \tilde{\Omega}_Q -1 )  - \Theta ( \tilde{\Omega}_Q - e^t )}{ 
 r_t ( Q ) }
 \nonumber
 \\
 & = &
 \frac{  
 \Theta ( e^t > \tilde{\Omega}_Q > 1 )}{ 
 r_t ( Q ) }
 \; ,
 \label{eq:Gtres}
\end{eqnarray}
where
 \begin{equation}
 \Theta ( x_2 > x > x_1 ) =
 \left\{
 \begin{array}{lc}
 1 & \mbox{if $ x_2 > x > x_1 $} \\
 0 & \mbox{else}
 \end{array}
 \right.
 \; ,
 \end{equation}
and 
 \begin{equation}
 \tilde{\Omega}_Q = \frac{{\Omega}_K}{\xi}
 \; .
 \end{equation}
Note that for the choice $\Omega_{K} = | \epsilon_{\bf{k}} -  
\epsilon_{ {\bf{k}}_F } | $ we obtain
$\tilde{\Omega}_Q  = | \xi^{\hat{\bf{n}} }_t ( q ) | \approx | q |$ 
to leading order.
The initial condition at $t=0$ implies
 \begin{eqnarray}
 r_{0} ( Q ) & = & \frac{Z_0^{\hat{\bf{n}}}}{\xi_0} 
\left[ i \omega 
 - \epsilon_{\bf{k}} + \mu - \Sigma_{\xi_0} ( K ) \right] 
 \nonumber
 \\
& = &
Z_0^{\hat{\bf{n}}} [ i \epsilon -  \xi_0^{\hat{\bf{n}}} ( q ) ]
 + \tilde{\Gamma}_{0}^{(2)} ( Q )
 \; .
 \end{eqnarray}
Furthermore, without interactions $Z_t = 1$ and 
$\tilde{\Gamma}_t^{(2)} ( Q ) = 0$, so that
$ r_{t} ( Q ) = i \epsilon -  \xi_t^{\hat{\bf{n}}} ( q)$.

\subsubsection{Higher order vertices ($n \geq 2$)}
The dimensionless scaling form of the higher order irreducible vertices 
follows from our definition (\ref{eq:fieldscale}) of the dimensionless fields,
\begin{eqnarray}
\tilde{\Gamma}_{t}^{ (2n) }
( Q_1^{\prime} , \ldots , Q_n^{\prime} ; Q_n , \ldots , Q_1 ) =
& & 
 \nonumber \\
& & \hspace{-40mm} \nu_0^{n-1} \xi^{n-2}
\left[ Z_t^{\hat{\bf{n}}_{1}^{\prime} } 
\cdots Z_t^{\hat{\bf{n}}_{n}^{\prime}}
 Z_t^{\hat{\bf{n}}_{n} } \cdots Z_t^{\hat{\bf{n}}_{1}} 
 \right]^{1/2}
 \nonumber
 \\
 & & \hspace{-40mm} \times
\Gamma^{(2n)}_{\xi} ( K_1^{\prime} , \ldots , K_n^{\prime} 
; K_n , \ldots ,
K_1 )
 \; .
 \label{eq:Gammarescale}
\end{eqnarray}

\subsection{Flow equations for the rescaled vertices}
\label{subsec:flowvertexrescale}

\subsubsection{Free energy}

Defining the interaction correction to 
the free energy per Fourier component,  
\begin{equation}
f_t = \frac{ \tilde{\Gamma}_{t}^{(0)} 
}{ \beta V \int_K \Theta ( \xi - \Omega_K ) }
\; ,
\end{equation}
and using Eqs.(\ref{eq:flowGamma0}) and (\ref{eq:Gamma0t}),
we obtain
the exact flow equation for $f_t$,
\begin{equation}
\partial_t f_t = f_t - \int \frac{ d S_{ \hat{\bf{n}}}}{S_D} 
 \eta_t^{\hat{\bf{n}}} -  \frac{  \int_Q
\delta ( \tilde{\Omega}_Q -1 ) \ln \left( \frac{ r_t 
( Q )}{ i \epsilon - \xi^{\hat{\bf{n}} }_t ( q )   } \right)
}{ \int_Q \Theta ( 1 - \tilde{\Omega}_Q ) }
 \; .
 \label{eq:fflow}
\end{equation}
Here $\eta_t^{\hat{\bf{n}}}$ is the flowing anomalous dimension,  
\begin{equation}
\eta_t^{\hat{\bf{n}}} = - \partial_t \ln Z_t^{\hat{\bf{n}}}
 = - \frac{\partial_t Z_t^{\hat{\bf{n}}} }{Z_t^{\hat{\bf{n}}} }
 \; .
\end{equation}
Eq.(\ref{eq:fflow}) is the fermionic analog of
the corresponding flow equation for the free energy in $\phi^4$-theory,
see Eq.(4.6) of Ref.\cite{Kopietz01}.

\subsubsection{Two-point vertex}
From Eqs.(\ref{eq:flowGamma2}) and (\ref{eq:Gammatdef}) we find
the flow equation for the dimensionless subtracted two-point vertex,
 \begin{eqnarray}
 \partial_t \tilde{\Gamma}_t^{(2)} ( Q ) & = &
 ( 1 - \eta_t^{\hat{\bf{n}}} - Q \cdot \partial_Q )
 \tilde{\Gamma}_t^{(2)} ( Q )  
 \nonumber
 \\
 & - &  \int_{Q^{\prime}}
 \dot{G}_t ( Q^{\prime} ) 
 \tilde{\Gamma}_{t}^{(4)}
 ( Q , Q^{\prime} ; Q^{\prime} , Q )
 \; ,
 \label{eq:twopointscale}
 \end{eqnarray}
where we have introduced the notation
 \begin{equation}
 Q \cdot \partial_Q 
 = q \partial_q + \epsilon \partial_{\epsilon}
 \; ,
 \end{equation}
 \begin{equation}
 \dot{{G}}_t  ( Q ) = \frac{ \delta ( \tilde{\Omega}_Q -1 )}{
 r_t ( Q )}
 \; .
 \end{equation}

\subsubsection{Four-point vertex}
The flow equation for the rescaled irreducible four-point vertex follows
from Eqs.(\ref{eq:flowGamma4}) and (\ref{eq:Gammarescale}), 
 \begin{eqnarray}
 \partial_t \tilde{\Gamma}_{t}^{ (4)}
 ( Q_1^{\prime} , Q_2^{\prime} ; Q_2 , Q_1 ) 
 & = & 
\nonumber
 \\
 &  & \hspace{-35mm}
 - \sum_{i = 1}^{2} 
 \left[  \frac{   \eta^{{\hat{\bf{n}}_{ i}^{\prime}}}_t + 
 \eta^{{\hat{\bf{n}}_{ i}}}_t }{2} +
 Q_i^{\prime} \cdot \partial_{ Q_i^{\prime}}  + 
 Q_i \cdot \partial_{ Q_i} 
 \right] 
 \tilde{\Gamma}_{t}^{(4) }
 ( Q_1^{\prime} , Q_2^{\prime} ; Q_2 , Q_1 )
 \nonumber
 \\
 &   & \hspace{-35mm} - \int_Q 
 \dot{G}_t ( Q )
  \tilde{\Gamma}_{t}^{(6)}
 ( Q_1^{\prime} , Q_2^{\prime} , Q ; Q , Q_2 , Q_1 )
 \nonumber
 \\
 &  & \hspace{-35mm} -  \int_Q 
 \left[ \dot{G}_t ( Q )  \tilde{G}_t 
( Q^{\prime} )  + 
 \tilde{G}_t ( Q )  \dot{G}_t 
 ( Q^{\prime} )
 \right]
 \nonumber
 \\
 &  & \hspace{-35mm} \times \left\{ \frac{1}{2}
 \left[
 \tilde{\Gamma}_{t}^{(4)}
 ( Q_1^{\prime} , Q_2^{\prime} ; Q^{\prime} , Q )
 \tilde{\Gamma}_{t}^{ (4) }
 ( Q , Q^{\prime} ; Q_2 , Q_1 )
 \right]_{ K^{\prime} = K_1 + K_2 - K}
 \right.
 \nonumber
 \\
 &  & 
 \hspace{-29mm} -  \left[
 \tilde{\Gamma}_{t}^{ (4) }
 ( Q_1^{\prime} , Q^{\prime} ; Q , Q_1 )
 \tilde{\Gamma}_{t}^{(4) }
 ( Q_2^{\prime} , Q ; Q^{\prime} , Q_2 )
 \right]_{ K^{\prime} = K + K_1 - K_1^{\prime}}
 \nonumber
 \\
 &  &
 \left.
 \hspace{-29mm}
 + 
\left[
 \tilde{\Gamma}_{t}^{(4) }
 ( Q_2^{\prime} , Q^{\prime} ; Q , Q_1 )
 \tilde{\Gamma}_{t}^{(4) }
 ( Q_1^{\prime} , Q ; Q^{\prime} , Q_2 )
 \right]_{ K^{\prime} = K + K_1 - K_2^{\prime}}
 \right\}
 \label{eq:fourpointscale}
 \; .
 \end{eqnarray}
Here $K$, $K^{\prime}$, $K_i$ and $K_i^{\prime}$ should be
considered as functions of the dimensionless scaling variables
introduced in Sec.\ref{subsec:scaling},
for example $ K = ( \hat{\bf{n}} k_F  + \hat{\bf{v}}_F \frac{\xi}{v_F} q , 
i \xi \epsilon )$. Due to the non-linearity of this transformation,
the explicit expression of $ Q^{\prime}$  
in terms of  $Q$
and  $Q_i = ( \hat{\bf{n}}_i , q_i , \epsilon_i)$ is 
rather complicated. 
For example, let us calculate 
$Q^{\prime} = (
\hat{\bf{n}}^{\prime}, q^{\prime} , \epsilon^{\prime} )$
in the zero-sound contribution involving
$K^{\prime} = K + K_1 - K_1^{\prime}$.
The energy component is simple,
$\epsilon^{\prime} = \epsilon + \epsilon_1 - \epsilon_1^{\prime}$,
but for $\hat{\bf{n}}^{\prime}$ and $q^{\prime}$ we obtain 
 \begin{equation}
 \hat{\bf{n}}^{\prime} =
 \frac{ \hat{\bf{n}}  + \hat{\bf{n}}_1 - \hat{\bf{n}}_1^{\prime} + c_t ( 
 \hat{\bf{n}} q  + \hat{\bf{n}}_1 q_1 - \hat{\bf{n}}_1^{\prime} q_1^{\prime} ) }{
 | \hat{\bf{n}}  + \hat{\bf{n}}_1 - \hat{\bf{n}}_1^{\prime} + c_t ( 
 \hat{\bf{n}} q  + \hat{\bf{n}}_1 q_1 - 
 \hat{\bf{n}}_1^{\prime} q_1^{\prime} ) |}
 \; ,
 \label{eq:hatnprime}
 \end{equation}
 \begin{equation}
 q^{\prime} = \frac{1}{c_t}
 \left[ \left|
 \hat{\bf{n}}  + \hat{\bf{n}}_1 - \hat{\bf{n}}_1^{\prime} + c_t ( 
 \hat{\bf{n}} q  + \hat{\bf{n}}_1 q_1 - \hat{\bf{n}}_1^{\prime} q_1^{\prime} )
 \right| -1 \right]
 \label{eq:qprimedef}
 \; ,
 \end{equation}
where $c_t = \xi_0 e^{-t} / ( m v_F^2)$, see Eq.(\ref{eq:ctdef}).
For simplicity we have assumed a spherical Fermi surface, so that
${\bf{k}}_i = \hat{\bf{n}}_i k_F ( 1 + c_t q_i)$.
Note that the corresponding rescaled energy dispersion is
 \begin{eqnarray}
 \xi^{\hat{\bf{n}}^{\prime}}_t ( q^{\prime})
 & = &
 q + \hat{\bf{n}} \cdot ( \hat{\bf{n}}_1 q_1 - \hat{\bf{n}}_1^{\prime}
 q_1^{\prime}) +  \frac{ \hat{\bf{n}} \cdot ( \hat{\bf{n}}_1 - 
 \hat{\bf{n}}_1^{\prime} )}{c_t}
 \nonumber
 \\
 & + &
\frac{c_t}{2} \left(  
 \hat{\bf{n}} q + \hat{\bf{n}}_1 q_1 - \hat{\bf{n}}_1^{\prime}
 q_1^{\prime}
 \right)^2 + \frac{( \hat{\bf{n}}_1 - 
 \hat{\bf{n}}_1^{\prime} )^2}{2 c_t}
 \nonumber
 \\
 & + & 
 ( \hat{\bf{n}}_1 - 
 \hat{\bf{n}}_1^{\prime} ) \cdot  
 ( \hat{\bf{n}} q +
 \hat{\bf{n}}_1 q_1  - \hat{\bf{n}}_1^{\prime} q_1^{\prime})
 \; .
 \label{eq:xiprime}
 \end{eqnarray}
The rescaled flow equation
for the six-point vertex is given in the Appendix.

\section{Advantages of rescaled flow equations}
\label{sec:advantages}
\setcounter{equation}{0}

In this section we discuss several properties  
of the rescaled flow equations of Sec.\ref{sec:rescale}, 
and argue that for practical calculations
it may be advantageous  to use the rescaled flow equations instead of
the unrescaled equations discussed in Sec.\ref{sec:mode}.  

\subsection{Screening and flow of the density-density correlation function
within RPA}

As pointed out in 
Refs.\cite{Halboth99,Halboth00,Honerkamp00,Honerkamp01},
the unrescaled flow equations predict pathological
RG flows for physical quantities which are
determined by degrees of freedom in the immediate vicinity
of the Fermi surface, such as  
uniform susceptibilities:
for any finite infrared cutoff $\xi$
uniform susceptibilities are not renormalized at all, 
while at $\xi = 0$ their RG flow exhibits a discontinuity.
Although for a Fermi liquid one may use
the Fermi liquid relations between the 
uniform susceptibilities and the quasi-particle interactions
to obtain flow equations for the susceptibilities
at finite cutoff $\xi$ \cite{Halboth99,Halboth00}, 
it would be better to calculate uniform susceptibilities
entirely within the framework of the exact RG, without
relying on the assumption that the system is a Fermi  liquid.

To illustrate the above point,
let us consider the screening of the effective interaction,
which is closely related to the 
density-density correlation function.
The screening problem has also been considered by
Shankar (see Appendix A of Ref.\cite{Shankar94}) within the
field theory version of the RG,
and by Dupuis \cite{Dupuis98} within the conventional Kadanoff-Wilson
RG scheme for Fermi liquids.
  
Consider the flow equation (\ref{eq:flowGamma4}) for the irreducible
four-point vertex. 
Suppose we start from a bare interaction
at scale $\xi_0$ of the form
 \begin{equation}
 \Gamma_{\xi_0}^{(4)} ( K_1^{\prime} , K_2^{\prime} ;
 K_2 , K_1 ) = f_{\xi_0} ( {K}_1 - K_1^{\prime} ) - f_{\xi_0}
 ( K_1 - K_2^{\prime} )
 \; .
 \label{eq:bare}
 \end{equation}
Let us now calculate the flow of the effective interaction
within the random-phase approximation (RPA), where the
BCS-contribution is ignored and only those zero-sound terms are
retained which preserve the form (\ref{eq:bare}). 
Setting for simplicity $K_1 = K_1^{\prime}$ and $K_2 = K_2^{\prime}$
(this is the combination appearing in the
flow equation (\ref{eq:flowGamma2}) for the self-energy)
we approximate the
irreducible four-point vertex at scale $\xi < \xi_0$ by
 \begin{equation}
 \Gamma_{\xi}^{(4)} ( K_1 , K_2 ;
 K_2 , K_1 ) \approx - f_{\xi} ( K_1 - K_2 )
 \; ,
 \label{eq:gamma4rpa}
 \end{equation}
where we have assumed $f_{\xi} ( 0 ) = 0$.
In this approximation we obtain from 
Eq.(\ref{eq:flowGamma4})
 \begin{equation}
  \partial_{\xi} f_{\xi}^{-1} (  K_1 - K_2 )  = \dot{\Pi}_{\xi} ( K_1 - K_2 )
  \; ,
 \label{eq:RPAflow}
 \end{equation}
where the 
change of the polarization is given by
 \begin{eqnarray}
  \dot{\Pi}_{\xi} ( K_1 - K_2 ) & = & \int_K \left[ 
 \frac{ \delta ( \Omega_K - \xi ) G_{\xi , \xi_0} ( K + K_1 - K_2 ) }{
 i \omega_n - \epsilon_{\bf{k}} + \mu - \Sigma_{\xi} ( K ) }
 \right.
 \nonumber
 \\
 & & \hspace{-21mm} \left.
 +
 \frac{ G_{\xi , \xi_0} ( K ) \delta ( \Omega_{K+ K_1 - K_2} - \xi ) }{
 i \omega_{n+n_1 - n_2} - \epsilon_{ {\bf{k}} + {\bf{k}}_1 - {\bf{k}}_2 } 
 + \mu - \Sigma_{\xi} ( K + K_1 - K_2  ) } \right]
 \; .
 \nonumber
 \\
 & &
 \label{eq:Pidot}
 \end{eqnarray}
Within the RPA we ignore the self-energy corrections
to the propagators in Eq.(\ref{eq:Pidot}). 
Using the cutoff function $\Omega_{K } = | \epsilon_{\bf{k}}
- \epsilon_{ {\bf{k}}_F } | $, we obtain for $| {\bf{k}}_1 - {\bf{k}}_2 | \ll k_F$
  \begin{eqnarray}
  \dot{\Pi}_{\xi} ( K_1 - K_2 ) & \approx & 2 \delta ( \xi )
 \int \frac{ d {\bf{k}} }{ ( 2 \pi )^D} \delta ( \epsilon_{\bf{k}} - 
\epsilon_{ {\bf{k}}_F } )
 \nonumber
 \\
 & \times &
 \frac{ {\bf{v}}_F \cdot ( {\bf{k}}_1 - {\bf{k}}_2 ) }{ 
 i ( \omega_{n_1} - \omega_{n_2})   - {\bf{v}}_F \cdot ( {\bf{k}}_1 - 
 {\bf{k}}_2) }
 \; .
 \label{eq:Pidotrpa}
 \end{eqnarray}
Obviously,  $\dot{\Pi}_{\xi} ( P )  = 0$ for any $\xi \neq 0$, so that
the RPA interaction $f_{\xi} ( P )$ is not renormalized
for any finite $\xi$. 
Due to the factor $\delta ( \xi )$ in Eq.(\ref{eq:Pidotrpa})
the RG flow of the RPA interaction $f_{\xi} ( P )$
(and hence  the flow of the polarization and the compressibility) is
discontinuous at $\xi = 0$. 
This discontinuity is smoothed out if one works at a finite
temperature \cite{Dupuis98,Honerkamp01b}, but  there
 is a way to avoid
discontinuous flow equations at zero temperature: 
the RG flow equations 
generated by our rescaled version of the exact RG remain
continuous even at $T =0$.

We now explicitly show  this for the RG-flow of the
rescaled density-density correlation function.
Within the RPA this the rescaled four-point vertex 
is approximated by
   \begin{equation}
  \tilde{\Gamma}_{t}^{(4)} ( Q_1 , Q_2 ;
 Q_2 , Q_1 ) \approx  - F_{t} ( Q_1 , Q_2 )
 \; ,
 \label{eq:tildeGamma4RPA}
 \end{equation}
where the dimensionless function $F_t ( Q_1 , Q_2 )$ satisfies
the flow equation
 \begin{equation}
 [ \partial_t - Q_1 \cdot \partial_{Q_1} 
 - Q_2 \cdot \partial_{Q_2} ] 
 F_t^{-1} ( Q_1 , Q_2 ) = - \dot{\tilde{\Pi}}_t ( Q_1 , Q_2 )
 \; ,
 \label{eq:flowcontinuous}
 \end{equation}
which is the rescaled analog of Eq.(\ref{eq:RPAflow}).
Here 
 \begin{equation}
 \dot{\tilde{\Pi}}_t ( Q_1 , Q_2 ) = \int_Q 
 \left[ \dot{G}_t ( Q ) \tilde{G}_t ( Q^{\prime} ) +
  \tilde{G}_t ( Q ) \dot{G}_t ( Q^{\prime} ) \right]_{K^{\prime}
 = K + K_1 - K_2}
 \; ,
 \label{eq:Pidot2}
 \end{equation}
where $Q^{\prime} = Q^{\prime} ( Q, Q_1, Q_2) $ 
is defined as a function of
$Q = ( \hat{\bf{n}}, q , i \epsilon) $, $Q_1 = ( \hat{\bf{n}}_1 , q_1 ,
i \epsilon_1 )$ and $Q_2  = ( \hat{\bf{n}}_2 , q_2 , i \epsilon_2)$ via
 $ K^{\prime} = K + K_1 - K_2$, as explained
in Sec.\ref{subsec:flowvertexrescale}.
Ignoring again interaction corrections to the propagators and
working with the  momentum shell $\tilde{\Omega}_Q = 
| \epsilon_{\bf{k}} - \epsilon_{ {\bf{k}}_F } | / \xi =
| \xi^{\hat{\bf{n}}}_t ( q ) |$,
we have
 \begin{equation}
 \tilde{G}_t ( Q )  = \frac{ \Theta ( e^t > 
  | \xi_t^{\hat{\bf{n}}} ( q ) | > 1 )}{ i 
 \epsilon -  \xi_t^{\hat{\bf{n}}} ( q ) }
 \; ,
 \end{equation}
 \begin{equation}
 \dot{G}_t ( Q )  = \frac{ \delta ( 
  | \xi_t^{\hat{\bf{n}}} ( q ) | - 1 )}{ i 
 \epsilon -  \xi_t^{\hat{\bf{n}}} ( q ) }
 \; .
 \end{equation}
Substituting these expressions into Eq.(\ref{eq:Pidot2})
and performing the integrations over $q$ and $\epsilon$ 
we obtain
 \begin{equation}
 \dot{\tilde{\Pi}}_t ( Q_1 , Q_2 ) = - 4 
 \int \frac{ d S_{\hat{\bf{n}}}}{S_D} \frac{ 
 \Theta ( 1 + e^t > 
  \hat{\bf{n}} \cdot {\bf{q}}_{12} >  2 )
 \hat{\bf{n}} \cdot {\bf{q}}_{12} }{
 ( \hat{\bf{n}} \cdot {\bf{q}}_{12} )^2 + ( \epsilon_1 - \epsilon_2 )^2 }
 \; ,
 \label{eq:Pidotscale}
 \end{equation}
where we have used Eq.(\ref{eq:xiprime})
to simplify  $\xi^{ \hat{\bf{n}}^{\prime} }_t ( {q^{\prime}})$
for small $ | {\bf{k}}_1 - {\bf{k}}_2 |$,
 \begin{equation}
 \xi^{ \hat{\bf{n}}^{\prime} }_t ( {q^{\prime}}) \approx q + \hat{\bf{n}} \cdot
 {\bf{q}}_{12} \; \; , \; \; {\bf{q}}_{12} = 
 {\hat{\bf{n}}}_1 q_1 - {\hat{\bf{n}}}_2 q_2 + \frac{ \hat{\bf{n}}_1 -
 \hat{\bf{n}}_2 }{c_t}
 \; .
 \end{equation}
The crucial point is now that, unlike Eq.(\ref{eq:Pidotrpa}),
the right-hand side of Eq.(\ref{eq:Pidotscale})
is a non-singular function of the flow-parameter $t$.
In fact, in dimensions $D > 1$ the $\Theta$-function discontinuity
of the integrand  disappears after the angular integration, so that
$ \dot{\tilde{\Pi}}_t ( Q_1 , Q_2 )$ is a continuous function of
$t$. But even in $D=1$, where the angular integration
should be replaced by a summation over the two discrete Fermi points,
there is no discontinuity in the solution of 
the flow equation (\ref{eq:flowcontinuous}).  

It is instructive to elaborate a little bit more on the case
$D=1$, where we obtain from Eq.(\ref{eq:Pidotscale}) 
 \begin{equation}
 \dot{\tilde{\Pi}}_t ( Q_1 , Q_2 ) = - 2
  \delta_{ \hat{\bf{n}}_1 , 
 \hat{\bf{n}}_2 } 
 \frac{ | q_1 - q_2 |
 \Theta ( 1 + e^t >
  | q_1 -q_2 | > 2   ) }{
 ( q_1 - q_2 )^2 + ( \epsilon_1 - \epsilon_2 )^2 }
 \; .
 \label{eq:Pidotscale1}
 \end{equation}
In deriving this expression we have assumed $c_t | q_i | \ll 1$, and
have  neglected the terms with
$\hat{\bf{n}}_1 = - \hat{\bf{n}}_2$, which
vanish exponentially for $t \rightarrow \infty$.
Note that 
the right-hand side of Eq.(\ref{eq:Pidotscale1}) 
has discontinuities, but no $\delta$-function
singularity, in contrast to Eq.(\ref{eq:Pidotrpa}).
 It follows that the flow 
of the physical density-density correlation function is continuous.
To see this, let us explicitly solve Eq.(\ref{eq:flowcontinuous}).
Using the method described in Ref.\cite{Kopietz01}, the solution 
with the correct initial condition  is easily obtained,
 \begin{eqnarray}
 F_t^{-1} ( Q_1 , Q_2 ) & = &  F_{t=0}^{-1} ( Q_1 , Q_2 )
 \nonumber
 \\
 &  & \hspace{-20mm} - \int_{0}^{t} d \tau 
 \dot{\tilde{\Pi}}_{t- \tau} ( {\hat{\bf{n}}}_1 , e^{- \tau} q_1 ,
 e^{- \tau } i \epsilon_1 ;   {\hat{\bf{n}}}_2 , e^{- \tau} q_2 ,
 e^{- \tau } i \epsilon_2 )
 \; .
 \nonumber
 \\
 & &
 \label{eq:Ftsolutionscale}
 \end{eqnarray}
Substituting Eq.(\ref{eq:Pidotscale1}) into Eq.(\ref{eq:Ftsolutionscale})
we finally obtain in $D = 1$
  \begin{eqnarray}
 F_t^{-1} ( Q_1 , Q_2 )  & = &  F_{t=0}^{-1} ( Q_1 , Q_2 )
 \nonumber
 \\ 
 & + &  \delta_{ \hat{\bf{n}}_1 , 
 \hat{\bf{n}}_2 } \tilde{\Pi}_{t} ( q_1 - q_2, i \epsilon_1 - i \epsilon_2 )
 \; ,
 \end{eqnarray}
where
 \begin{eqnarray}
 \tilde{\Pi}_{t} ( q, i \epsilon )
 & = & \frac{ | q |}{ q^2 + \epsilon^2} \left[
 ( | q | - 2 ) \Theta ( 1 + e^t > | q | > 2    )
 \right.
 \nonumber
 \\
 & + & 
 \left.
 ( 2 e^t - | q | ) \Theta (  2 e^t >  | q | > 1 +  e^{t} ) \right]
 \label{eq:Piflow1res}
 \; .
 \end{eqnarray}
Note that the function $x \Theta ( x )$ is continuous, so that
the right-hand side of Eq.(\ref{eq:Piflow1res})
is indeed a continuous function of $t$.
The physical density-density correlation function is
then
 \begin{eqnarray}
 \Pi_{\xi} ( p , i \omega ) & = & 
 \nu_0 \tilde{\Pi}_t \left( \frac{v_F p}{\xi} ,
 \frac{ i \omega }{\xi} \right)
 \nonumber
 \\
 &  & \hspace{-10mm} = \frac{ 1}{\pi } 
 \left[
 \frac{ | p |  ( v_F | p | - 2 \xi )
 \Theta (  \xi + \xi_0 > v_F | p | > 2 \xi  ) }{
 ( v_F p )^2 + \omega^2 }
 \right. 
 \nonumber
 \\
 & & 
 \left. \hspace{-7mm} +
 \frac{ | p |
 ( 2 \xi_0 - v_F | p |  )
 \Theta (  2 \xi_0   > v_F | p | > \xi + \xi_0 )}{
 ( v_F p )^2 + \omega^2 }
 \right]
 \label{eq:unrescale}
 \; .
 \end{eqnarray}
Taking the limits $ \xi_0 \rightarrow \infty$
and $\xi \rightarrow 0$ we obtain the well-known result
 \begin{equation}
 {\Pi}_{0} ( p, i \omega ) = \frac{v_F}{\pi } \frac{ p^2}{( v_F p )^2 + 
\omega^2 }
 \; .
 \end{equation}
From the first line in Eq.(\ref{eq:unrescale}) it is clear
why our rescaled flow equations are more suitable for the calculation
of the susceptibilities than the corresponding unrescaled flow equations:
Because the dependence of the density-density correlation function
on the infrared cutoff appears in the scaling function
$ \tilde{\Pi}_t \left( \frac{v_F p}{\xi} ,
 \frac{ i \omega }{\xi} \right)$
via the ratios $p / \xi$ and $\omega / \xi$,
the result for uniform susceptibilities depends on the 
order in which the limits $ p \rightarrow 0$ and $\xi \rightarrow 0$ are 
taken. Our rescaled RG equation directly 
yields the {\it{scaling function}}
$ \tilde{\Pi}_t \left( \frac{v_F p}{\xi} ,
 \frac{ i \omega }{\xi} \right)$
where this problem does not arise.
Let us give an alternative explanation for this difference:
in the unrescaled flow equations the momenta and frequencies are
held constant as the infrared cutoff is reduced. Any fixed distance
from the Fermi surface is therefore magnified on the scale of the reduced
infrared cutoff,
so that momenta that are
initially close to the Fermi surface 
are mapped onto new momenta which, on the reduced scale, appear 
further away from the Fermi surface.
On the other hand,
in the rescaled flow equations the  momenta and 
frequencies are scaled down together 
with the infrared cutoff, so that
the  degrees of freedom that are initially in the vicinity of the Fermi surface
are mapped onto coarse grained degrees of freedom
that remain close to the Fermi surface.

\subsection{The Fermi surface as a RG fixed point manifold}
\label{sec:FSshape}

As already emphasized in Sec.\ref{subsec:FS},
to obtain well-behaved scaling properties 
we should expand the bare energy dispersion 
around the true Fermi surface of the interacting many-body problem,
which a priori is not known. 
We now show how the shape of the
Fermi surface can be calculated a posteriori
from the fixed point equation for the momentum- and 
frequency-independent part of the 
irreducible two-point vertex,
 \begin{equation}
 \tilde{\mu}_t^{\hat{\bf{n}}} = \tilde{\Gamma}_t^{(2)} 
( {\hat{\bf{n}}} ,  0 , i 0)
 \label{eq:mutildedef}
 \; .
 \end{equation}
Note that this is a relevant coupling function
(labeled by the direction $\hat{\bf{n}}$) with scaling dimension
$+1$. There are 
two marginal coupling functions associated with the two-point vertex.
One of them is the usual
wave-function renormalization given in Eq.(\ref{eq:Zdef}). 
In terms of our rescaled two-point vertex it can also be
written as
 \begin{equation}
 Z_t^{\hat{\bf{n}}} = 1 - \left. \frac{\partial 
 \tilde{\Gamma}^{(2)}_t (\hat{\bf{n}} ,  q=0 , i \epsilon )}{\partial
 ( i \epsilon ) } \right|_{\epsilon=0}
 \; .
 \end{equation}
The other marginal coupling can be taken to be 
the dimensionless renormalization factor of the Fermi velocity
 \begin{equation}
 \tilde{v}_t^{\hat{\bf{n}}} = Z_t^{\hat{\bf{n}}} - 
 \left. \frac{\partial 
 \tilde{\Gamma}_t^{(2)} ( {\hat{\bf{n}}} , q , i 0 )}{\partial
 q } \right|_{q=0}
 \; .
 \end{equation}
The RG flow of the shape of the Fermi surface
of strongly correlated electrons in reduced dimensions
have recently been considered by several authors 
\cite{Shankar94,Halboth00,Honerkamp00,Louis01}.
However, the interpretation of the renormalized Fermi surface
in terms of a RG fixed point manifold 
has only been emphasized by Shankar \cite{Shankar94}, who 
used the conventional
one-loop momentum shell technique, including the rescaling step.
In contrast, the authors of Ref.\cite{Halboth00,Honerkamp00,Louis01}
work with unrescaled flow equations, in which case
the coupling corresponding to Eq.(\ref{eq:mutildedef}) 
does not flow to a RG fixed point, so that
the interpretation of the Fermi surface as a RG fixed point
manifold is obscured.

To study the shape of the Fermi surface within the exact RG,
let us retain only relevant and marginal couplings in the
two-point function. In this approximation
 \begin{equation}
 r_t ( Q ) 
 \approx  i \epsilon - \tilde{v}_t^{\hat{\bf{n}}} q +
 {\tilde{\mu}}_t^{\hat{\bf{n}}}  
 \; .
 \label{eq:rtapprox}
 \end{equation}
It is convenient to define the 
Fermi surface momenta ${\bf{k}}_{F,t}$
of the system with  cutoff
$\xi = \xi_0 e^{-t}$ in analogy with Eq.(\ref{eq:FSdef}),
 \begin{equation}
 \epsilon_{ {\bf{k}}_{F,t}} = \mu - \Sigma_{ \xi} (
 {\bf{k}}_{F,t} , i 0 )
 \; ,
 \label{eq:FSflow}
 \end{equation}
so that the true
renormalized Fermi surface is 
$ {\bf{k}}_F = \lim_{ t \rightarrow \infty} {\bf{k}}_{F, t}$,
see  Eq.(\ref{eq:FSdef}).
In order to calculate $ {\bf{k}}_{F,t}$ we use the fact that
by construction $q$ in Eq.(\ref{eq:rtapprox}) 
is measured relative to the
renormalized ${\bf{k}}_F$, so that we obtain from
the definition (\ref{eq:FSflow}), up to irrelevant terms,
 \begin{equation}
 {\bf{k}}_{F,t} = {\bf{k}}_{F }
 + {\hat{\bf{v}}}_{F }  \frac{  \xi_0 e^{-t}
   \tilde{\mu}_{t}^{\hat{\bf{n}}}  }{ v_{F } 
 \tilde{v}^{\hat{\bf{n}}}_t }
 \; .
 \label{eq:kfren}
 \end{equation}
Setting $t =0$ we 
obtain for the difference
between the fully renormalized Fermi momentum
${\bf{k}}_F$
and the Fermi momentum ${\bf{k}}_{F,0}$ of the model with 
cutoff $\xi_0$,
 \begin{equation}
 {\bf{k}}_{F} - {\bf{k}}_{F,0} = - \hat{\bf{v}}_{F} 
 \frac{ \xi_0  \tilde{\mu}_0^{\hat{\bf{n}}}  }{v_{F} 
 \tilde{v}_0^{\hat{\bf{n}}}}
 \; .
 \end{equation}
It should be kept in mind that ${\bf{v}}_F$ is defined in terms of the
gradient of the {\it{bare}} energy dispersion at the 
{\it{renormalized}} ${\bf{k}}_F$.
From  Eqs.(\ref{eq:Gammatdef}) and
(\ref{eq:mutildedef}) we find 
 \begin{equation}
 \xi_0 \tilde{\mu}_0^{\hat{\bf{n}}} = 
 Z_0^{\hat{\bf{n}}} 
 [ \Sigma ( {\bf{k}}_F , i 0 ) - \Sigma_{\xi_0} ( {\bf{k}}_F , i 0) ]
 \; ,
 \label{eq:countermu}
 \end{equation}
which allows us to reconstruct the counterterm
$\Sigma ( {\bf{k}}_F , i 0 )$ as follows:
Suppose that we have adjusted the initial condition
$\tilde{\mu}_0^{\hat{\bf{n}}}$
such that for $t \rightarrow \infty$ the relevant couplings
$\tilde{\mu}_t^{\hat{\bf{n}}}$
approach a fixed point. 
Due to the relevance of
$\tilde{\mu}_t^{\hat{\bf{n}}}$, we expect
that this requires some fine tuning of the initial 
$\tilde{\mu}_0^{\hat{\bf{n}}}$, so that
on the critical manifold
(i.e. the manifold in parameter space that flows into a
RG fixed point) 
$\tilde{\mu}_0^{\hat{\bf{n}}}$ becomes a function 
of the other relevant and marginal couplings.
From Eq.(\ref{eq:countermu}) we then obtain
the counterterm $\Sigma ( {\bf{k}}_F , i 0 )$
as a function of  these couplings.
Note that if we adjust the initial conditions such that
$\tilde{\mu}^{\hat{\bf{n}}}_t$ remains finite for $t \rightarrow \infty$,
then Eq.(\ref{eq:kfren}) guarantees
that for $ t \rightarrow \infty$ the flowing
${\bf{k}}_{F,t}$ indeed approaches the true Fermi surface.
{\it{Thus, the problem of constructing the renormalized Fermi surface 
can be reduced to the problem of 
finding the RG fixed point of the
flow equation for the relevant coupling function
${\tilde{\mu}}_t^{\hat{\bf{n}}}$.}}

The above discussion relies on the 
approximation  (\ref{eq:rtapprox}) for the inverse propagator,
which is only justified
for a Fermi liquid. 
In principle it is also possible that the system flows to a
non-Fermi liquid
fixed point, characterized by some new renormalized Fermi surface. 
As pointed out by Anderson \cite{Anderson93}, in this case there
might be some subtleties related to the 
subtraction of the counterterm  
$\Sigma ( {\bf{k}}_F , i 0 )$.

It is also worth emphasizing the following point:
In order to reach a RG fixed point, it is crucial that
the bare energy dispersion is expanded around
the fully {\it{renormalized}} Fermi surface ${\bf{k}}_F$, defined 
in Eq.(\ref{eq:FSdef}). 
If we had chosen the bare
Fermi surface ${\bf{k}}_{F,0}$ as the reference for the expansion,
then the RG flow of the relevant 
coupling $\tilde{\mu}_t^{\hat{\bf{n}}}$ would 
exhibit a runaway flow, i.e. 
$| \tilde{\mu}_t^{\hat{\bf{n}}} | \rightarrow \infty$ for
$t \rightarrow \infty$.
To see this, suppose that
we expand around the initial Fermi surface at scale $\xi_0$. Then 
the relation (\ref{eq:kfren}) between the 
renormalized and the bare Fermi momentum is replaced by 
 \begin{equation}
 {\bf{k}}_{F,t} = {\bf{k}}_{F,0 }
  + {\hat{\bf{v}}}_{F,0 }  \frac{  \xi_0 e^{-t}
    \tilde{\mu}_{t}^{\hat{\bf{n}}}   }{ v_{F,0 } 
 \tilde{v}^{\hat{\bf{n}}}_t }
 \; ,
 \label{eq:kfren2}
 \end{equation}
where ${\bf{v}}_{F,0} = \left. \nabla_{\bf{k}} \epsilon_{\bf{k}}
 \right|_{ {\bf{k}}_{F,0} }$.
Given that ${\bf{k}}_{F,t}$ approaches for large $t$  a value that is
different from the bare ${\bf{k}}_{F,0}$, we see from
Eq.(\ref{eq:kfren2}) that
$ \tilde{\mu}_{t}^{\hat{\bf{n}}}$ must necessarily diverge as
$e^{t}$  (assuming that $ \tilde{v}^{\hat{\bf{n}}}_t$ remains finite).
This runaway flow indicates that we have expanded around the
wrong Fermi surface. This phenomenon is well known 
from the usual theory of critical phenomena, where
the fixed point manifold is discrete:
In this case a runaway flow indicates the existence of some new 
fixed point.
What is new here is that the fixed point manifold is a continuum, which can
therefore continuously change under RG transformations. 
To avoid a runaway flow, we have to expand all quantities from the beginning
around the true Fermi surface, which can be determined a posteriori
from the requirement that the RG flow indeed approaches a fixed point.

\subsection{Rescaling and the possibility of
non-Fermi liquid fixed points}
\label{subsec:wilson}

Long time ago Bell and Wilson \cite{Bell74} pointed out
that the field rescaling (see step 3 in Sec.\ref{sec:intro})
is necessary in order
to obtain a non-trivial RG fixed point, characterized by
a non-zero anomalous dimension $\eta$. 
In fact, the RG fixed point conditions can be viewed as some sort
of non-linear eigenvalue equations, which  contain $\eta$ as an adjustable
parameter \cite{Comellas98}. 
From this point of view it is not surprising \cite{Bagnuls00} that
solutions of the fixed point equations 
exist only for certain values of $\eta$, depending on the interaction, the
dimensionality, and the symmetries of the system.
We believe that the above statement is 
also true for fermionic many-body systems,
so that the unrescaled flow equations used in 
Refs.\cite{Zanchi96,Halboth99,Halboth00,Honerkamp00,Honerkamp01}
cannot be used to detect non-Fermi liquid fixed points.

Because within a one-loop approximation the anomalous dimension 
vanishes, at the first sight it seems that 
at this level of approximation 
it is sufficient to work with a pure mode-elimination 
RG \cite{Zanchi96,Halboth99,Halboth00,Honerkamp00,Honerkamp01}.
This is not necessarily true, however, because
even at the one-loop level the rescaling of momenta and frequencies
(see step 2 in Sec.\ref{sec:intro})
can be essential to obtain a non-trivial RG fixed point.
For example, consider
$\phi^{4}$-theory slightly below  $D=4$ dimensions.
In this case there are two relevant couplings, namely the
momentum-independent part of the irreducible two-point vertex
$\tilde{\mu}_t$, and 
the momentum-independent part of the irreducible four-point vertex
$\tilde{g}_t$.
The one-loop RG flow equations 
for suitably defined \cite{Kopietz01} dimensionless couplings
are well known \cite{ZinnJustin89}
 \begin{eqnarray}
 \partial_t \tilde{\mu}_t  & = & 2 \tilde{\mu}_t + 
\frac{ \tilde{g}_t }{2 ( 1 +
 \tilde{\mu}_t ) }
 \; , 
 \label{eq:muflow}
 \\
 \partial_t \tilde{g}_t & = & ( 4 -D ) \tilde{g}_t - 
  \frac{ 3 \tilde{g}_t^2}{ 2 (1 + \tilde{\mu}_t )^2 }
 \; .
 \label{eq:gflow}
 \end{eqnarray}
For $ 0 < 4 - D \ll 1$ these equations have a non-trivial fixed point at
 \begin{equation}
 \tilde{g}^{\ast} \approx \frac{2}{3} ( 4 - D )
 \; \; , \; \; \tilde{\mu}^{\ast} \approx - \frac{1}{6} ( 4 - D )
 \; ,
 \end{equation}
which arises from a
{\it{balance between the competing effects}} due to the
scaling terms
(i.e. the first terms on the right-hand sides
of  Eqs.(\ref{eq:muflow}) and (\ref{eq:gflow}))
and the mode elimination terms
(i.e. the second terms on the right-hand sides
of  Eqs.(\ref{eq:muflow}) and (\ref{eq:gflow})).
Obviously, a RG transformation which includes 
only the mode elimination 
would miss the Wilson-Fisher fixed point.

Finally, we point out that the rescaling of
momenta and frequencies plays also an important role to
accelerate  the flow of the irrelevant couplings to negligibly small values.
For example, 
the constant part of the rescaled irreducible six-point vertex  
$\tilde{\Gamma}^{(6)}_t ( Q_1^{\prime} , Q_2^{\prime} , 
Q_3^{\prime} ; Q_3 , Q_2 , Q_1 )$ is irrelevant with scaling dimension
$-1$. This $-1$ appears as the first term
on the right-hand side of Eq.(\ref{eq:sixpointscale}), and
causes the size of the rescaled six-point vertex
to  diminish exponentially as we  iterate the RG
(as long as this decrease is not overwhelmed by some anomalous dimension).
On the other hand, the corresponding
flow equation (\ref{eq:sixpoint}) for the unrescaled six-point vertex 
does not exhibit a similar 
suppression due to a negative scaling dimension, so that 
there is no guarantee that the unrescaled six-point vertex 
really becomes small as we iterate the RG transformation.

\section{Summary and conclusions}
\setcounter{equation}{0}

In this work we have constructed exact functional RG equations
for non-relativistic fermions, taking not only the
mode elimination into account, but also 
the rescaling of momenta, frequencies, and fields.
The exact flow equations given in this work
are valid in arbitrary dimensions and
describe the RG flow of any translationally invariant
fermionic many-body system in the normal state.
We emphasize that these equations 
are exact and describe the RG flow
of the irreducible correlation functions of the many-body system
for all momenta and frequencies. Of course,
it is possible to extract the
usual perturbative RG $\beta$-function
describing the flow of
the momentum-independent part of the four-point vertex
from these flow equations.
To obtain the $\beta$-function at the two-loop
order, it is sufficient to truncate the infinite hierarchy
of flow equations by setting $\Gamma^{(2n)} = 0$ for $ n \geq 4$.
In Sections \ref{subsec:sharp}, \ref{subsec:flowvertexrescale} and in 
the Appendix
we have explicitly written down all expressions
necessary for performing a two-loop calculation for any
$D$-dimensional
translationally invariant normal fermionic many-body system.
These equations are also valid in $D=1$, where
two-loop calculations are usually performed using the field
theory-method \cite{Solyom79}. Note, however, that
in this case there are some subtleties 
which are  not fully understood \cite{Carta00}.

We have formulated the
exact functional RG for fermionic systems
in close analogy with the functional RG for
field theories and statistical mechanics problems.
We have tried to emphasize this analogy by
adopting as much as possible the notation used
in statistical mechanics \cite{Wegner73,Kopietz01}.
A crucial step is the introduction of dimensionless scaling variables
and rescaled vertices, which describe the scaling
towards the true Fermi surface of the system.
Finally, in Sec.\ref{sec:advantages}
we have shown that the
 rescaled flow equations for the irreducible
vertices given in Sec.\ref{subsec:flowvertexrescale}
have several advantages as compared with the unrescaled 
flow equations discussed in Sec.\ref{subsec:sharp}:

\begin{enumerate}

\item {\it{Flow of susceptibilities.}}
The unrescaled exact RG equations used in Refs.
\cite{Zanchi96,Halboth99,Halboth00,Honerkamp00,Honerkamp01}
yield discontinuities in flow equations
for quantities which are dominated by degrees of freedom in the
immediate vicinity of the Fermi surface, such as the
density-density correlation function at long wavelengths.
In contrast, our formulation of the exact RG including rescaling
yields
flow equations which are continuous functions of the flow parameter.

\item {\it{Renormalization of the shape of the Fermi surface.}}
The calculation of the shape of the renormalized
Fermi surface can be reduced to the solution of the
fixed point equation for the 
properly rescaled irreducible two-point vertex at vanishing
external momentum and frequency.
Without rescaling, the corresponding flow equation does not 
have a fixed point, so that the associated relevant coupling
exhibits a runaway flow.

\item {\it{Non-Fermi liquid fixed points.}}
If the system has a scale invariant non-Fermi liquid fixed point
characterized by a non-zero anomalous dimension $\eta$, 
then the flow generated by RG transformations which omit 
the rescaling steps 2 and 3 listed in Sec.\ref{sec:intro}
cannot detect this fixed point.
Instead,
the pure mode elimination RG calculations of the type
used in Refs.\cite{Zanchi96,Halboth99,Halboth00,Honerkamp00,Honerkamp01}
would generate a runaway-flow to strong coupling in this case.

\end{enumerate}

We suspect that our rescaled flow equations 
will also turn out to be advantageous for numerical calculations, because
they describe the RG flow of the relevant scaling functions.
The numerical analysis of these equations is a rather tedious 
task \cite{Halboth99,Honerkamp00}, which is beyond
the scope of this work. 

Our flow equations also offer a new approach to
study interacting fermions in $D =1$, where the
normal metallic state is known to be 
a non-Fermi liquid.
For these systems the RG $\beta$-function is often calculated within the
field theory formulation of the RG advanced by 
S\'{o}lyom \cite{Solyom79}. 
However, till now there exists no RG calculation
of the full momentum- and frequency-dependent
single-particle Green function
of interacting fermions in  $D=1$.
From bosonization it is known
that in momentum-frequency space
the single-particle Green function 
exhibits interesting features such as power-law singularities,
which are a manifestation of 
anomalous scaling and spin-charge-separation \cite{Voit94}.
We are currently  investigating whether the full spectral function
of interacting fermions in $D=1$ can be calculated
by means of an approximate solution of the rescaled functional RG equation
(\ref{eq:twopointscale}) for the two-point vertex.
We have preliminary evidence \cite{Busche01} that at weak coupling
this can indeed be done, and that for the Tomonaga-Luttinger model
the resulting spectral function compares quite well with
the exact result obtained via bosonization.

Finally, we would like to address the question whether 
possible scale-invariant
non-Fermi liquid fixed points
in dimensions $D > 1$ are accessible within the exact RG. 
Certainly, detecting such a fixed point requires at least a two-loop calculation,
because at the one-loop level the wave-function renormalization
$Z_t^{\hat{\bf{n}}}$ remains unity. Considering the complexity of numerical analysis of the
one-loop equations describing the RG flow without 
the rescaling steps \cite{Halboth99,Halboth00,Honerkamp00,Honerkamp01},
the direct numerical analysis  of the two-loop RG flow equations 
including rescaling seems to be a rather difficult task.
Such a calculation would require the numerical analysis of the flow equation
for the six-point vertex given in Eq. (\ref{eq:sixpointscale}).
Note, however, that in the special case 
of a square Fermi surface 
Binz, Baeriswyl and Dou\c{c}ot \cite{Binz01} have recently 
presented an analytic study of the 
one-loop RG flow in the vicinity of the dominant instabilities.
Let us emphasize again that $Z_t^{\hat{\bf{n}}} \rightarrow 0$ at a non-Fermi liquid fixed point,
so that  the inclusion of the flow of the wave-function
renormalization is crucial to reach such a fixed point. 

For Fermi surfaces with a special geometry it is possible
to analyze the RG flow analytically \cite{Binz01,Ferraz01}.
In particular, very recently Ferraz \cite{Ferraz01} 
performed a field-theoretic two-loop RG calculation for a two-dimensional 
Fermi system with a truncated Fermi surface, consisting of flat and curved
pieces.  He succeeded to calculate the entire single-particle spectral function.
Interestingly, he found non-Fermi liquid behavior for all points on the
Fermi surface, including the curved pieces.
The calculations presented in  Ref. \cite{Ferraz01}  are quite encouraging and
support our point of view that
higher-dimensional non-Fermi liquid fixed points
are accessible  within 
a two-loop truncation of the Wilsonian RG equations presented in this work.

\section*{Acknowledgments}
We thank B. Binz, A. Ferraz, C. Honerkamp, 
V. Meden, M. Salmhofer, and K. Sch\"{o}nhammer
for discussions, and N. Dupuis for pointing out some
relevant references.
This work was financially supported by the DFG-Schwerpunkt
SSP 1053, project No.SCHO201/8-1.

\appendix
\renewcommand{\theequation}{A.\arabic{equation}}
\renewcommand{\thesubsection}{A.\arabic{subsection}}

\section*{Flow equations for the irreducible six-point vertex}

\begin{figure}
\epsfysize9cm 
\hspace{5mm}
\epsfbox{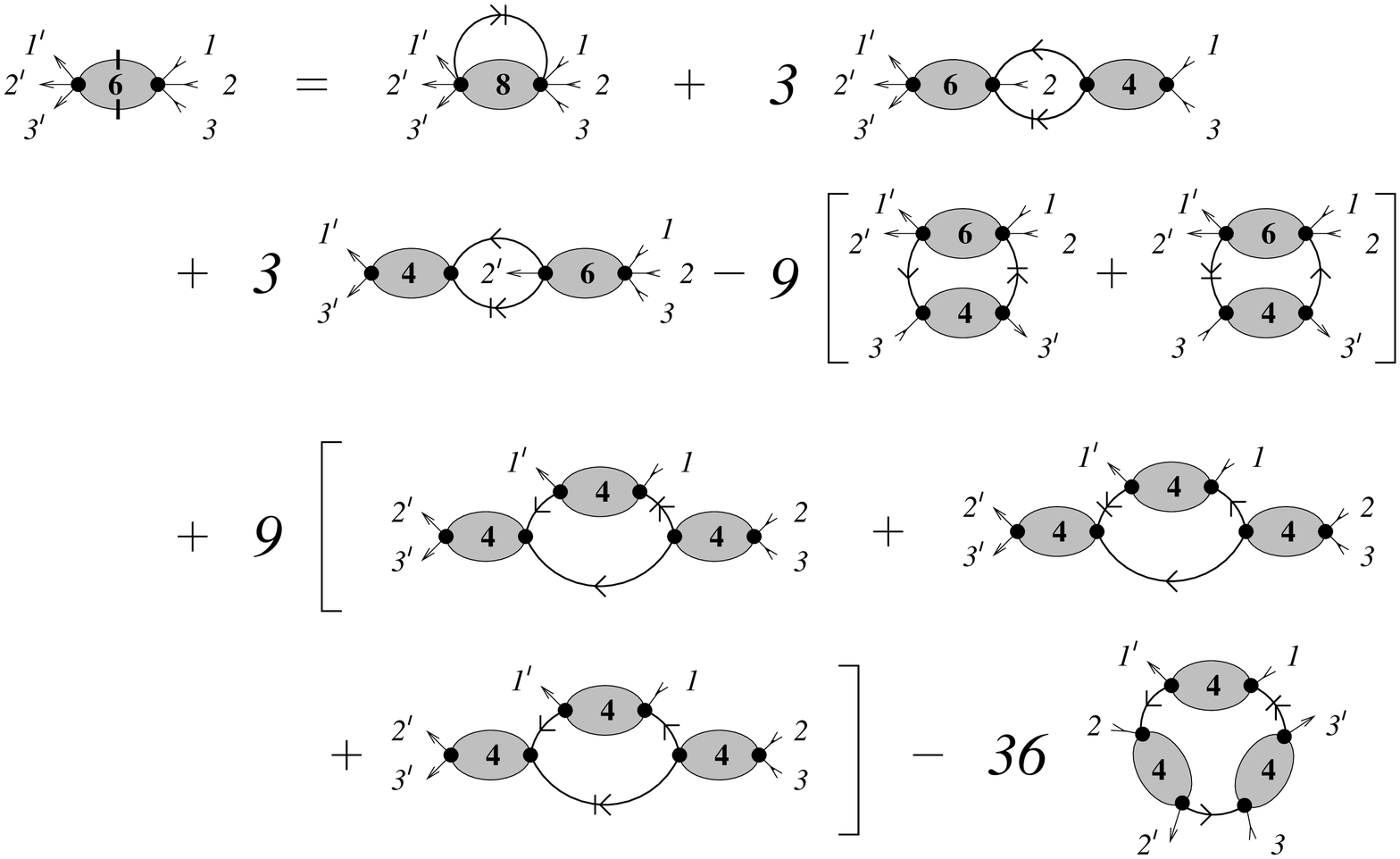}
\vspace{5mm}
\caption{
Diagrammatic representation of the flow equation
for the six-point vertex, see Eqs.(\ref{eq:sixpoint}) and
(\ref{eq:sixpointscale}).
}
\label{fig:SixPoint}
\end{figure}

For a two-loop calculation we need 
the flow equation for the irreducible six-point 
vertex \cite{Morris94,Kopietz01,Busche01}.
The unrescaled flow equation reads 
(see Fig.\ref{fig:SixPoint} and recall that for fermions $\zeta = -1$) 
 {\small
 \begin{eqnarray}
 \partial_\xi \Gamma^{(6)}_{\xi }
 ( K_1^{\prime} , K_2^{\prime} , K_3^{\prime} ; 
 K_3 , K_2  , K_1 ) 
 & = & 
 \nonumber
 \\
 &   & \hspace{-53mm} - \zeta \int_K
  \frac{ \delta ( \Omega_K - \xi )}{
 i \omega_{n} - \epsilon_{ {\bf{k}} } + \mu
 - \Sigma_{\xi} ( K ) }
 \Gamma^{(8)}_{ \xi }
 ( K_1^{\prime} , K_2^{\prime} , K_3^{\prime} , 
 K ; K , K_3 ,  K_2 , K_1 )
 \nonumber
 \\
 & & \hspace{-53mm } + 3 \int_K
 \frac{ \delta ( \Omega_K - \xi )   G_{\xi , \xi_0} ( K^{\prime} )  }{
 i \omega_{n} - \epsilon_{ {\bf{k}} } + \mu
 - \Sigma_{\xi} ( K ) }
 \nonumber
 \\
 & & \hspace{-43mm}
 \times  \left\{
 {\cal{A}}_{(1^{\prime},2^{\prime}),3^{\prime}}
 \left[
 \Gamma^{(4)}_{ \xi }
 ( K_1^{\prime} , K_2^{\prime} ; K^{\prime} , K )
 \Gamma^{(6)}_{ \xi }
 ( K , K^{\prime} , K_3^{\prime} ; K_3 , K_2 , K_1 )
 \right]_{ K^{\prime} = K_1^{\prime} + K_2^{\prime} - K}
 \right.
 \nonumber
 \\
 &  & 
 \hspace{-38mm} \left. + {\cal{A}}_{1, ( 3, 2 )}
  \left[
 \Gamma^{(6)}_{ \xi }
 ( K_1^{\prime} , K_2^{\prime}, K_3^{\prime} ; 
 K^{\prime} , K , K_1 )
 \Gamma^{(4)}_{ \xi }
 ( K , K^{\prime} ; K_3 , K_2 )
 \right]_{ K^{\prime} = K_2 + K_3 - K}
 \right\} 
 \nonumber
 \\
 & & \hspace{-53mm } + 9 \zeta \int_K
 \left[  \frac{ \delta ( \Omega_K - \xi )
 G_{\xi , \xi_0} ( K^{\prime} )}{
 i \omega_{n} - \epsilon_{ {\bf{k}} } + \mu
 - \Sigma_{\xi} ( K ) }
 +
 \frac{  G_{\xi , \xi_0} ( K )   
 \delta ( \Omega_{K^{\prime}} - \xi )}{
 i \omega_{n^{\prime}} - \epsilon_{ {\bf{k}}^{\prime} } + \mu
 - \Sigma_{\xi} ( K^{\prime} ) } \right]
 \nonumber
 \\
 & & \hspace{-43mm}
 \times  
 {\cal{A}}_{(1^{\prime},2^{\prime}),3^{\prime}} 
 {\cal{A}}_{(2,1), 3 }
 \left[
 \Gamma^{(6)}_{\xi }
 ( K_1^{\prime} , K_2^{\prime}, K^{\prime} ; K , K_2 , K_1 )
 \Gamma^{(4)}_{ \xi }
 ( K_3^{\prime} , K ; K^{\prime} , K_3 )
 \right]_{ K^{\prime} = K + K_3^{\prime} - K_3 }
 \nonumber
 \\
  & & \hspace{-53mm } + 9  \int_K 
 \left[  \frac{ \delta ( \Omega_K - \xi ) 
 G_{\xi , \xi_0} ( K^{\prime} )
 G_{\xi , \xi_0} ( K^{\prime \prime} ) }{
 i \omega_{n} - \epsilon_{ {\bf{k}} } + \mu
 - \Sigma_{\xi} ( K ) } 
 +
 \frac{   G_{\xi , \xi_0} ( K )    
 \delta (  \Omega_{K^{\prime}}  - \xi )
 G_{\xi , \xi_0} ( K^{\prime \prime} ) }{
 i \omega_{n^{\prime}} - \epsilon_{ {\bf{k}}^{\prime} } + \mu
 - \Sigma_{\xi} ( K^{\prime} ) } 
 \right.
 \nonumber
 \\
 & & \hspace{-43mm} \left.
 +
 \frac{  G_{\xi , \xi_0} ( K ) G_{\xi , \xi_0} ( K^{\prime } )
 \delta (  \Omega_{K^{\prime \prime}} - \xi )}{
 i \omega_{n^{\prime \prime}} - \epsilon_{ {\bf{k}}^{\prime \prime} } + \mu
 - \Sigma_{\xi} ( K^{\prime \prime} ) } 
\right] {\cal{A}}_{(1^{\prime},2^{\prime}),3^{\prime}} 
 {\cal{A}}_{3, ( 2, 1 )}
 \nonumber
 \\
 & & \hspace{-41mm}
 \times  
 \left[
 \Gamma^{(4)}_{ \xi  }
 ( K_1^{\prime} , K_2^{\prime} ; K , K^{\prime} )
 \Gamma^{(4)}_{\xi }
 ( K_3^{\prime} , K^{\prime} ; K^{\prime \prime} , K_3 )
 \Gamma^{(4)}_{ \xi }
 ( K^{\prime \prime} , K^{\prime} ; K_2 , K_1 )
 \right]_{ K^{\prime} = K_1 + K_2
 + K_3 - K_3^{\prime} - K}^{ K^{\prime \prime} = K_2 + 
 K_1 - K}
 \nonumber
 \\
  & & \hspace{-53mm } - 36 \zeta  \int_K 
 \frac{ \delta ( \Omega_K - \xi ) 
 G_{\xi , \xi_0} ( K^{\prime} )
 G_{\xi , \xi_0} ( K^{\prime \prime} ) }{
 i \omega_{n} - \epsilon_{ {\bf{k}} } + \mu
 - \Sigma_{\xi} ( K ) } 
 {\cal{A}}_{1^{\prime},  2^{\prime}, 3^{\prime} }
 {\cal{A}}_{1,2,3} 
 \nonumber
 \\
 & & \hspace{-43mm}
 \times   
 \left[
 \Gamma^{(4)}_{ \xi  }
 ( K_1^{\prime} , K^{\prime} ; K , K_1 )
 \Gamma^{(4)}_{\xi  }
 ( K_2^{\prime} , K^{\prime \prime} ; K^{\prime} , K_2 )
 \Gamma^{(4)}_{\xi }
 ( K_3^{\prime} , K ; K^{\prime \prime} , K_3 )
 \right]_{ K^{\prime} = K_2^{\prime} + K_3^{\prime}
 + K  - K_3 - K_2}^{ K^{\prime \prime} = K_3^{\prime} + 
 K - K_3}
 \; .
 \nonumber
 \\
 & & 
 \label{eq:sixpoint}
 \end{eqnarray}
 }
Here the anti-symmetrization operators 
${\cal{A}}_{1,2,3}$ and ${\cal{A}}_{ 1,(2,3) }$ are defined as follows,
 \begin{eqnarray}
 {\cal{A}}_{ 1,2,3} f ( 1,2,3 ) & = & 
 \frac{1}{6} \left[ f ( 1,2,3) + f ( 2,3,1) + f (3,1,2) 
 \right.
 \nonumber
 \\
 & & \hspace{4mm} \left.
- f (3,2,1) -
 f ( 2,1,3 ) - f ( 1,3,2 ) \right]
\; ,
 \label{eq:A123}
 \\
  {\cal{A}}_{ 1,(2,3)} f ( 1,2,3 ) & =&
 {\cal{A}}_{ (2,3), 1} f ( 1,2,3 )
  = 
 \frac{1}{3} \left[ f ( 1,2,3) - f ( 2,1,3) - f (3,2,1) \right]
 \label{eq:A231}
 \; .
 \end{eqnarray}
Given a function $f (1,2,3)$ that is already 
antisymmetric with respect to the pair $(2,3)$, the function
${\cal{A}}_{ 1,(2,3)} f (1,2,3)$ 
is a totally antisymmetric function.
 Note that the combinatorial factors in front
of the terms involving the combinations $\Gamma^{(4)} \Gamma^{(6)}$ and
$\Gamma^{(4)} \Gamma^{(4)} \Gamma^{(4)}$ are
precisely the same as in the corresponding flow equation
of $\phi^4$-theory, see Eq.(4.20) of Ref.\cite{Kopietz01}.

The rescaled version of the above flow equation is
(setting now $\zeta = -1$)
 {\small
 \begin{eqnarray}
 \partial_t \tilde{\Gamma}_{t}^{(6) }
 ( Q_1^{\prime} , Q_2^{\prime} , Q_3^{\prime} ; 
 Q_3 , Q_2  , Q_1 ) 
 & = & 
 \nonumber
 \\
&  & \hspace{-50mm} 
 \left\{  -1 - \sum_{i = 1}^{3} 
 \left[ \frac{  \eta^{{\hat{\bf{n}}_{ i}^{\prime}}}_t + 
 \eta^{{\hat{\bf{n}}_{ i}}}_t }{2} +
 Q_i^{\prime} 
 \cdot \partial_{Q_i^{\prime}} + Q_i \cdot \partial_{Q_i} 
 \right] \right\} 
 \tilde{\Gamma}_{t}^{(6) }
 ( Q_1^{\prime} , Q_2^{\prime} , Q_3^{\prime} ; 
 Q_3 , Q_2  , Q_1 ) 
 \nonumber
 \\
 &   & \hspace{-50mm} - \int_Q 
 \dot{G}_t ( Q ) 
  \tilde{\Gamma}_{t}^{(8) }
 ( Q_1^{\prime} , Q_2^{\prime} , Q_3^{\prime} , 
 Q ; Q , Q_3 ,  Q_2 , Q_1 )
 \nonumber
 \\
 & & \hspace{-50mm } - 3  \int_Q   
 \dot{G}_t ( Q ) 
 \tilde{G}_t ( Q^{\prime} ) 
\left\{
 {\cal{A}}_{(1^{\prime},2^{\prime}),3^{\prime}}
 \left[
 \tilde{\Gamma}_{t}^{ (4) }
 ( Q_1^{\prime} , Q_2^{\prime} ; Q^{\prime} , Q )
 \tilde{\Gamma}_{t}^{(6) }
 ( Q , Q^{\prime} , Q_3^{\prime} ; Q_3 , Q_2 , Q_1 )
 \right]_{ K^{\prime} = K_1^{\prime} + K_2^{\prime} - K}
 \right.
 \nonumber
 \\
 &  & 
 \hspace{-41mm} \left. + {\cal{A}}_{1, ( 3, 2 )}
  \left[
 \tilde{\Gamma}_{t}^{(6) }
 ( Q_1^{\prime} , Q_2^{\prime}, Q_3^{\prime} ; 
 Q^{\prime} , Q , Q_1 )
 \tilde{\Gamma}_{t}^{(4) }
 ( Q , Q^{\prime} ; Q_3 , Q_2 )
 \right]_{ K^{\prime} = K_2 + K_3 - K}
 \right\} 
 \nonumber
 \\
 & & \hspace{-50mm } + 9  \int_Q 
 \left[ \dot{G}_t ( Q ) 
 \tilde{G}_t ( Q^{\prime} )  +
 \tilde{G}_t ( Q ) 
 \dot{G}_t ( Q^{\prime} ) \right]
 \nonumber
 \\
 & & \hspace{-44mm}
 \times  
 {\cal{A}}_{(1^{\prime},2^{\prime}),3^{\prime}} 
 {\cal{A}}_{(2,1), 3 }
 \left[
 \tilde{\Gamma}_{t}^{(6) }
 ( Q_1^{\prime} , Q_2^{\prime}, Q^{\prime} ; Q , Q_2 , Q_1 )
 \tilde{\Gamma}_{t}^{(4) }
 ( Q_3^{\prime} , Q ; Q^{\prime} , Q_3 )
 \right]_{ K^{\prime} = K + K_3^{\prime} - K_3 }
 \nonumber
 \\
  & & \hspace{-50mm } + 9 \int_Q 
 \left[
 \dot{G}_t ( Q ) 
 \tilde{G}_t ( Q^{\prime} ) 
 \tilde{G}_t ( Q^{\prime \prime} )
 +
 \tilde{G}_t ( Q ) 
 \dot{G}_t  ( Q^{\prime} ) 
 \tilde{G}_t ( Q^{\prime \prime} )
 +
  \tilde{G}_t ( Q ) 
 \tilde{G}_t ( Q^{\prime} ) 
 \dot{G}_t ( Q^{\prime \prime} )
  \right] {\cal{A}}_{(1^{\prime},2^{\prime}),3^{\prime}} 
 {\cal{A}}_{3, ( 2, 1 )}
 \nonumber
 \\
 & & \hspace{-44mm}
 \times  
 \left[
 \tilde{\Gamma}_{t}^{(4)  }
 ( Q_1^{\prime} , Q_2^{\prime} ; Q , Q^{\prime} )
 \tilde{\Gamma}_{t}^{(4) }
 ( Q_3^{\prime} , Q^{\prime} ; Q^{\prime \prime} , Q_3 )
 \tilde{\Gamma}_{t}^{ (4) }
 ( Q^{\prime \prime} , Q^{\prime} ; Q_2 , Q_1 )
 \right]_{ K^{\prime} = K_1 + K_2
 + K_3 - K_3^{\prime} - K}^{ K^{\prime \prime} = K_2 + 
 K_1 - K}
 \nonumber
 \\
  & & \hspace{-50mm } - 36  \int_Q 
 \dot{G}_t ( Q ) 
 \tilde{G}_t ( Q^{\prime} ) 
 \tilde{G}_t ( Q^{\prime \prime} ) 
 {\cal{A}}_{1^{\prime},  2^{\prime}, 3^{\prime} }
 {\cal{A}}_{1,2,3} 
 \nonumber
 \\
 & & \hspace{-44mm}
 \times   
 \left[
\tilde{\Gamma}_{t}^{(4) }
 ( Q_1^{\prime} , Q^{\prime} ; Q , Q_1 )
\tilde{ \Gamma}_{t}^{(4) }
 ( Q_2^{\prime} , Q^{\prime \prime} ; Q^{\prime} , Q_2 )
 \tilde{\Gamma}_{t}^{(4)  }
 ( Q_3^{\prime} , Q ; Q^{\prime \prime} , Q_3 )
 \right]_{ K^{\prime} = K_2^{\prime} + K_3^{\prime}
 + K  - K_3 - K_2}^{ K^{\prime \prime} = K_3^{\prime} + 
 K - K_3}
 \; .
 \label{eq:sixpointscale}
 \end{eqnarray}
 }

%

%


\begin{thebibliography}{99}
\bibitem{footnotegrass}
We introduce $\zeta = -1$
because then the bosonic version of the exact flow equations of
Sec.\ref{sec:mode}  can be simply obtained by replacing
$\zeta \rightarrow +1$.
\bibitem{footnoteflow}
Although we use a different notation, our Eq.(\ref{eq:flowgeneral})
is equivalent with 
the flow equation for the Legendre effective action
given in Refs.\cite{Honerkamp00,Honerkamp01}.
We have adopted the same notation as Ref.\cite{Kopietz01} to emphasize
the analogy with $\phi^4$-theory.
\end{thebibliography}
\end{document}